\documentclass{aa}  
\usepackage[version=4]{mhchem}
\usepackage{graphicx}
\usepackage{natbib}
\usepackage[breaklinks=true]{hyperref}
\hypersetup{linkcolor=red,citecolor=blue,filecolor=cyan,urlcolor=magenta}
\bibpunct{(}{)}{;}{a}{}{,}             

\newcommand{\cii}{[C\,{\sc ii}] }

\newcommand{\azir}{$73.4\pm2$}

\usepackage{txfonts}
\begin{document}

   \title{
   Exploring the environment, magnetic fields, and feedback effects of massive high-redshift galaxies with \cii}

   \subtitle{}
   \titlerunning{The environment, magnetic fields, and feedback of high-redshift massive galaxies}

   \author{K. Kade
          \inst{1}
          \and
          K.K. Knudsen
          \inst{1}
          \and 
          W. Vlemmings
          \inst{1}
          \and
          F. Stanley 
          \inst{2}
          \and
          B. Gullberg
          \inst{3}
          \and 
          S. König
          \inst{1}
          }

   \institute{Department of Space, Earth, \& Environment, Chalmers University of Technology, Chalmersplatsen 4 412 96 Gothenburg, Sweden \\
              \email{kiana.kade@chalmers.se} 
              \and
              Institut de Radioastronomie Millimétrique (IRAM), 300 Rue de la Piscine, 38400 Saint-Martin-d’Hères, France 
              \and 
              Cosmic Dawn Center DTU Space, Technical University of Denmark, Elektrovej 327, DK-2800 Kgs. Lyngby, Denmark}

   \date{Received July 21, 2021 / Accepted February, 2023}

  \abstract
   {Massive galaxies are expected to grow through different transformative evolutionary phases. High-redshift starburst galaxies and quasars are thought to be such phases and thus provide insight into galaxy evolution. Several physical mechanisms are predicted to play an important role in driving these phases; for example, interaction with companion galaxies, active galactic nuclei feedback, and possibly magnetic fields.
}
   {Our aim is to characterize the physical properties and the environment of the submillimeter galaxy AzTEC-3 at $z = 5.3$ and the lensed quasar BRI\,0952$-0115$ at $z = 4.4$, and to set a limit on the polarization properties of the two sources. We intend to place these two sources in the broader context of galaxy evolution, specifically star formation and mass growth through cosmic time.}
   {We used full polarization, sub-arcsecond-resolution, ALMA band-7 observations of both BRI\,0952$-0115$ and AzTEC-3. We detect \cii(\ce{{}^{2}P_{3/2}-{}^{2}P_{1/2}}) line emission towards both BRI\,0952$-0115$ and AzTEC-3, along with companions in each field. 
   We present an updated gravitational lensing model for BRI\,0952$-$0115 for correction of gravitational magnification. 
   }
   {We present infrared luminosities, star-formation rates, and \cii line to infrared luminosity ratios for each source. The \cii emission line profile for both BRI\,0952$-$0115 and AzTEC-3 exhibit a broad, complex morphology, indicating the possible presence of outflows. We present evidence of a "gas bridge" between AzTEC-3 and a companion source. 
   Modified blackbody spectral energy distribution fitting is used to analyze the properties of \cii detected companion sources in the field of both the submillimeter galaxy and the quasar. We investigated the possible role of the detected companions in outflow signatures. Using a simple dynamical mass estimate for the sources, we suggest that both systems are undergoing minor or major mergers. 
   No polarization is detected for the \cii, placing an upper limit below that of theoretical predictions.} 
   {Our results show that high-velocity wings are detected, indicating possible signs of massive outflows; however, the presence of companion galaxies can affect the final interpretation. Furthermore, the results provide additional evidence in support of the hypothesis that massive galaxies form in overdense regions, growing through minor or major mergers with companion sources. Finally, strong, ordered magnetic fields are unlikely to exist at the kiloparsec scale in the two studied sources.
   }


   \maketitle
%
\section{Introduction}

The discovery of intense starbursts ($100-1000 \rm M_{\odot}$) at high redshift demonstrates the prevalence of short, transformative phases in the evolution of massive galaxies \citep[e.g.,][]{Smail97,Hughes98,Blain99,Casey14}. Furthermore, the discovery of the relation between the mass of super-massive black holes (SMBHs) in the center of local massive galaxies and the velocity dispersion of those galaxies suggests coeval evolution between the SMBH and the host galaxy \citep[e.g.,][]{Magorrian98,Ferrarese00,Gebhardt00,Haering04,Gultekin09,Beifiori12,Kormendy13,Bennert15,Reines15}. The cosmic black hole accretion rate density has been found to follow a similar trend with redshift to that of the cosmic star-formation rate (SFR) density, indicating that a significant part of the evolution takes place during the first few billion years after the big bang (e.g., \citealt{Kormendy13} and \citealt{Madau14}). Thus, characterizing the physical properties of high-redshift starburst galaxies and quasars is essential for establishing a complete and coherent description of the evolution of massive galaxies. 

\begin{figure*}[t]
    \centering
    \includegraphics[width = 1.0\linewidth]{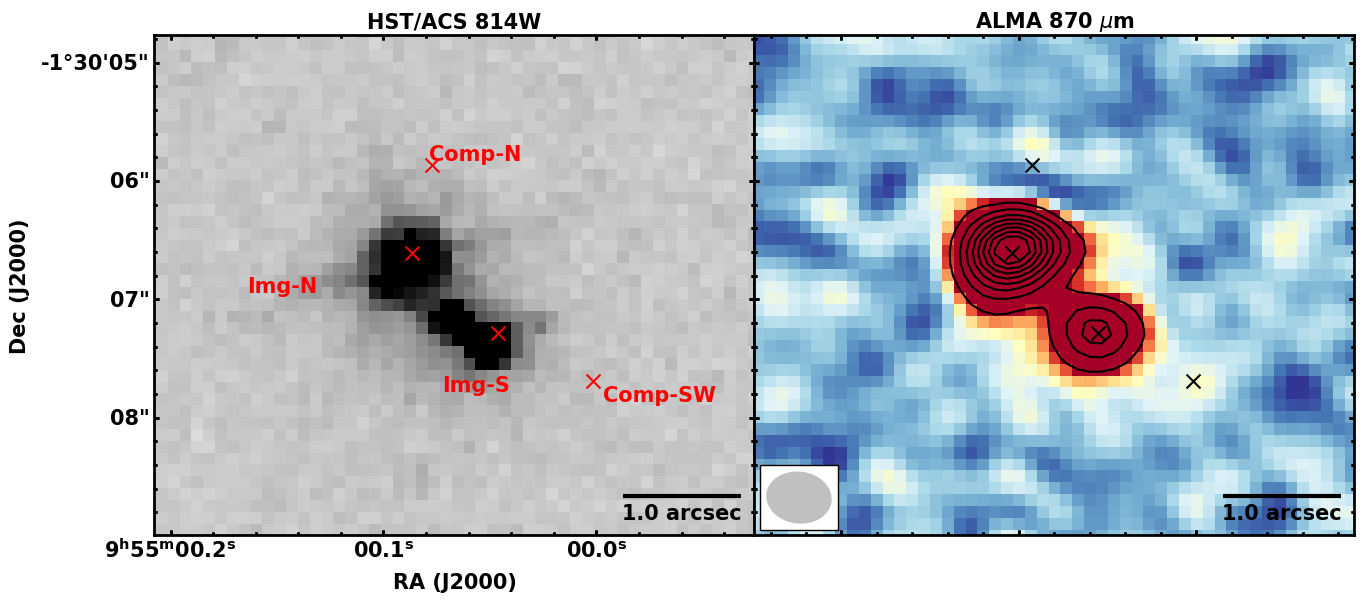}
    \caption{{\it Hubble Space Telescope (HST)}/ACS F814W (left) image of BRI0952 showing both lensed images of the quasar; Img-N and Img-S correspond to the two images of the quasar BRI0952 and Comp-N and Comp-SW to the companion sources. ALMA $870\,\mu$m continuum map (right) was created using line-free channels. The contours are shown at $5,10,20,30,40,50,60,70,$ and $80\,\sigma$ levels. The synthesized beam is shown in the bottom left of the ALMA image.}
    \label{fig:bri_info}
\end{figure*}

Many physical mechanisms play a role in massive galaxy evolution. Galaxies are thought to grow through two main mechanisms, namely galaxy mergers and gas accretion from the intergalactic medium \citep[e.g.,][]{Keres05,Hopkins08,Dekel09,Genzel10,Krumholz10,DiMatteo12,Dubois12}. Additionally, the presence of a growing SMBH, as seen in active galactic nuclei (AGNs) and quasars has been found in theoretical modeling to provide a mechanism for regulating star formation \citep[e.g.,][]{DiMatteo05,Narayanan15,Harrison17}. This implies that characterizing the environment of high-redshift quasars and starbursts, as well as understanding the role of AGN feedback, is paramount to understanding the manner in which these galaxies evolve.

Simulations and theoretical predictions from the early universe have demonstrated the likelihood of ubiquitous major and minor mergers occurring at high redshift \citep[e.g.,][]{Kaviraj13, Kaviraj15, Fogasy17}. Cosmological hydrodynamical simulations such as Illustris \citep{Sijacki15} and Horizon-AGN \citep{Dubois14} can shed light on the effects of these interactions. Studies have shown that minor mergers (classified as those with mass-ratios of >1:4) and the effect of companion galaxy interactions systematically affect the SFRs and evolution of massive starbursting galaxies \citep[e.g.,][]{Kaviraj15, Sparre16, Pearson19, Patton20}. 

Previous studies using optical, near-infrared imaging, or Lyman-$\alpha$ emission in the environment of high-redshift starburst galaxies and quasars have yielded conflicting results. Some of these systems show little or no evidence of companion sources \citep[e.g.,][]{Willott05,Banados13,Mazzucchelli17,Yue19}, while others find overdensities \citep[e.g.,][]{Carilli13_2,Husband15,Fan16}. However, in the last decade, an increasing number of companion sources around high-redshift galaxies have been discovered with the Atacama Large Millimeter/Submillimeter Array (ALMA), primarily using \cii observations \citep[e.g.,][]{Oteo16,trakhtenbrot17,Decarli17,DiazSantos18,wardlow18,Casey19,Jones19,Litke19,Neeleman19,Fogasy20,Venemans20}. 
\citet{DiazSantos18} studied the hot dust-obscured galaxy \footnote{Hot dust-obscured galaxies are a class of high-redshift, dust-obscured, AGN-dominated galaxies.}(Hot DOG) W2246$-$0526 and found evidence of a "gas-bridge" structure between a companion source and the central source, suggesting interaction-induced gas flow in the system. The system BRI\,1202$-$0725 has also been found to host multiple companions to the quasar and exhibits a similar bridge structure \citep{Carilli13_2}. 

Studies of the effect of AGNs have demonstrated that galactic-scale outflows and feedback can control star formation activity and black hole growth \citep[e.g.,][]{Fabian12, King15, Ishibashi16}. The extent of this phenomena is unknown, but recent studies of both low- and high-redshift galaxies show many of these sources exhibit outflow characteristics \citep[e.g.,][]{Heckman90, Rupke05, Weiner09, Banerji11, Cicone14, Chisholm15, Spilker20} through, for example, blueshifted absorption lines, P-Cygni line profiles, or broad emission line components typically seen at higher absolute velocities away from a main emission line. However, these findings become increasingly rare at higher redshifts due to the difficulty in detecting and determining outflow signatures in high-redshift galaxies. Although direct observations of individual sources with outflows are still rare at $z>4$ \citep[e.g.,][]{Maiolino12, Cicone15, Spilker20, Butler21}, studies using stacking of the \cii line have provided an alternative method of searching for outflows in a wider variety and number of sources, though also providing conflicting results \citep[e.g.,][]{Gallerani18, Decarli18, Bischetti19, Stanley19, Ginolfi20}. 

The strength of the \cii line facilitates its ubiquitous use at high redshift. At high redshifts, this line is shifted to the millimeter (mm) or submillimeter (sub-mm) regime where it is observable by ground-based facilities such as ALMA. The [CII] (\ce{^{2}P_{3/2}} $\rightarrow$ \ce{^{2}P_{1/2}}) emission is produced primarily in photodissociation regions (PDRs) by gas exposed to ultraviolet (UV) radiation and acts as one of the major coolants in star-forming regions of the interstellar medium (ISM) \citep[e.g.,][]{Stacey91,Stacey10,Carilli13}. For this reason, it has long been utilized to study the ISM of high-redshift galaxies.

\begin{figure*}
    \centering
    \includegraphics[width = 0.9\linewidth]{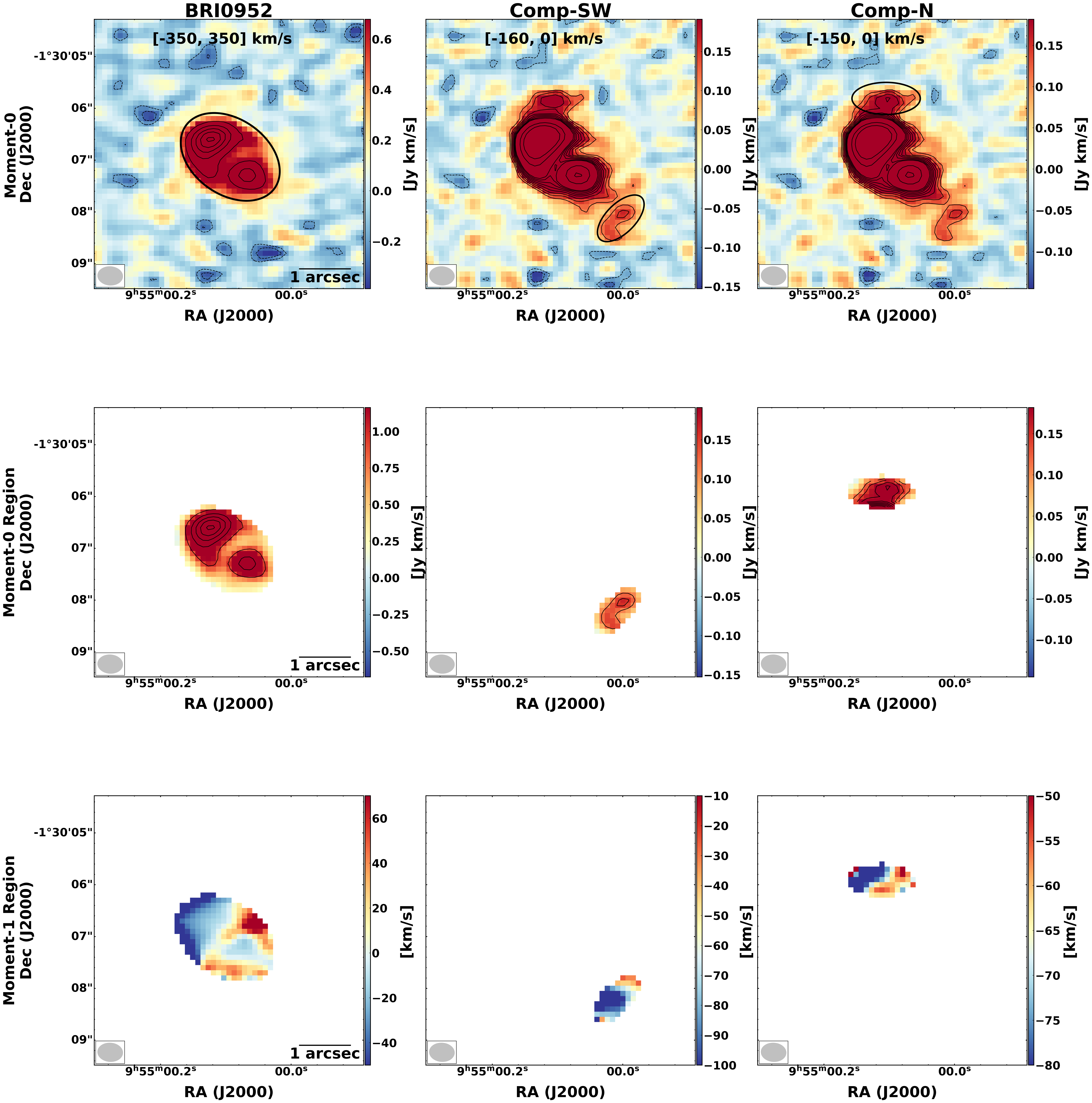}
    \caption{Moment-0 and moment-1 maps for the BRI0952 system. The top row shows moment-0 maps of the original \cii\! with velocity ranges for the individual extraction shown on the image. The middle row shows the moment-0 map of the region from which the spectra was extracted for the companions. The bottom row shows the moment-1 maps for the sources taken from the same region the spectra was extracted from. The contours are shown at $-3,-2,10,20,30,40,50,$ and $60\,\sigma$ levels for BRI0952, and contours are at $-3,-2,3,4,5,6,$ and $7\,\sigma$ levels for the companions where the respective $1\sigma$ noise level has been taken from individual moment-0 maps due to the restrictive velocity ranges (and thus differing noise levels). The black ellipses in the top row correspond to the region from which the spectra was extracted. The synthesized beam is shown in the bottom left corner of the images.}
    \label{fig:BRI_mom0}
\end{figure*}

Although magnetic fields are common, their presence and role in
galaxy formation and evolution in the early Universe is still
unclear. In the local Universe, ordered magnetic fields with a strength
of several $\mu$G are revealed from synchrotron and Faraday rotation
observations \citep[e.g.,][]{Beck2015} in normal spiral galaxies. In local
starburst galaxies, fields as high as $\sim20$~mG have been measured
\citep{Robishaw2008}, and ordered magnetic fields have been found
using infrared (IR) dust polarization observations \citep[e.g.,][]{Lopez2021b}. Exactly when
these fields were generated is unclear, but models show that a strong
regular field can quickly be generated from an initial weak seed field
as a result of turbulence driven by, for instance supernova explosions \citep{Rieder2017}. Since atoms and ions can easily be aligned by radiation and are subsequently very sensitive to realignment under the influence of only a weak magnetic field, the strength of the \cii line has led to the suggestion that \cii could be an excellent tracer of magnetic fields \citep{Yan2006, Zhang2015}.

In order to better constrain massive galaxy evolution at high redshift, high-sensitivity data is required to find faint emission from companion sources and to resolve the detailed physics ongoing within these systems. In this study, we explored the environment, searched for magnetic-field signatures, and examined possible outflow properties of two galaxies in the early Universe: the quasar BRI\,0952-0115 (hereafter BRI0952) at $z=4.433$ and the massive submillimeter galaxy (SMG) AzTEC-3 at $z=5.3,$ utilizing high-resolution ALMA band-7 \cii observations. These data were obtained through a proposal designed to look for magnetic fields in the early Universe and were selected based on their previously known extremely bright \cii emission. Due to the nature of the original project, the sources were observed in full polarization mode in an attempt to detect polarized emission from these sources. 

The AzTEC-3 protocluster has been extensively studied previously and encompasses the SMG itself (AzTEC-3) along with a quasar 13\,Mpc away in projected distance and a number of previously reported Lyman-break galaxies (LBGs) surrounding the SMG \citep{Capak11, Riechers14}. BRI0952 is a gravitationally lensed quasar and has previously been studied due to its strong \cii emission and lensing features \citep{Maiolino09, Gallerani12}. Here, we present findings of resolved \cii emission in both sources, along with an analysis of their surroundings. These findings help shed light on the role both sources have to play in the evolution of massive galaxies in the early Universe. 

In Section \ref{sec:obs}, we describe the ALMA observations and analysis utilized in this paper. In Section \ref{sec:results}, we describe the results from the \cii line and continuum analysis and the polarization of both AzTEC-3 and BRI0952. Section \ref{sec:discussion} provides a discussion of the properties of both sources as well as the environment these sources reside in. Finally, we present our conclusions in Section \ref{sec:conclusions}. Throughout this paper, we adopt the term "outflows" to describe the high-velocity flow of gas away from central regions typically associated with energy-driven winds by central AGNs or through starburst activity. Similarly, the term "gas-bridge" is used to describe structures of gas connecting components within galaxy systems thought to be caused by physical interactions between these components. Furthermore, we utilized a flat $\Lambda$CDM cosmology with $H_{0}$ = 70 km\,s$^{-1}$\,Mpc$^{-1}$,  $\Omega_{\rm M} = 0.3$, and $\Omega_{\Lambda}= 0.7$. 


\section{Observations and analysis} \label{sec:obs}
We obtained ALMA band-7 data for BRI0952 and AzTEC-3 as part of a program originally designed to search for magnetic field lines in the early Universe (2018.1.01536.S, P.I. Vlemmings). The setup for both galaxies was tuned to the \cii 1900.5369 GHz line, and the "time domain mode" was used. Observation details are listed in Table \ref{tab:obs}.  

The data calibration steps were performed following the ALMA polarization calibration scripts using CASA 5.4.0. (CASA \footnote{http://casa.nrao.edu/ \citep[CASA][]{mcmullin07}}). 
This includes calibration of the phase, bandpass, flux, and gain. 
The phase calibrators used for BRI0952 were J0948+002, J0854+2006, and J0725-0054, and for AzTEC-3 these were J0854-2006 and J0948+0022. The quasar J0854+2006 was used for polarization calibration. The uncertainty on the absolute flux calibration is conservatively estimated to be 10\%. 

\begin{table*}[h]
    \centering
    \caption{ALMA band-7 observation details for BRI0952 and AzTEC-3. We note that these correspond to images created using a robust factor of 0.0. }
    \begin{tabular}{l c c c c c} \hline \hline
        Source & Date & $N_{\rm ant}$ & $\nu_{\mathrm{spw, central}}$ & Synthesized Beam & Image RMS\\
         & [yyyy mm dd] &  [GHz] & [\,$''\times''$\,] & [mJy/beam] & [km/s] \\ \hline 
        BRI0952$_{\rm [CII]}$ & 2019 04 06 & 41  & 349.8 & 0.48 $\times$ 0.35 & 0.5\\
        BRI0952$_{\rm Cont}$ & & & 336.0,338.0,348.0 & $0.53 \times 0.41$ & 0.05 \\
        AzTEC-3 & 2019 03 22 & 48 & 301.7 & 0.80 $\times$ 0.56 & 0.2 \\ 
        AzTEC-3$_{\rm Cont}$ & & & 288.0,289.9,299.9 & $0.91 \times 0.65$ & 0.02 \\ \hline
        
    \end{tabular}
    \label{tab:obs}
\end{table*}

\begin{table}[h!]
    \caption{Linear polarization $3\sigma$-limits.}
    \label{polres}
    \centering
    \begin{tabular}{l l l l}
    \hline \hline
    Source & $P_{l, {\rm peak}}$ (\cii) & $P_{l, {\rm int}}$ (\cii) & $P_{l, {\rm dust}}$ \\
    \hline
    BRI0952 & $<4.8\%$ & $<0.7\%$ & $<2.7\%$\\
    AzTEC-3 & $<2.1\%$ & $<0.2\%$ & $<2.1\%$ \\
    \hline
    \end{tabular}
\end{table}

To analyze the Stokes-$I$ data, imaging was done using the task {\sc tclean}. All sources in the field were masked during cleaning. An initial search for line emission was done using "dirty" images to identify the line-free channels available for use in continuum subtraction, which was subsequently performed using the {\sc uvcontsub} task with a polynomial fit of the order of one for both AzTEC-3 and BRI0952 (see Figures \ref{fig:Az3_contsub} and \ref{fig:BRI_contsub}). The continuum subtraction for BRI0952 was complicated by the appearance of unexpected lines in spectral windows adjacent to that of the \cii line. \footnote{We note that the results of this paper do not change when using other tools such as {\sc imcontsub} or {\sc statcont} \citep{statcont18}.}

In the case of AzTEC-3, we used the frequency range of the observations up to 300\,GHz due to the possible additional \cii wing feature seen in the 300.5-301\,GHz range and the width of the \cii line (see Figures \ref{fig:Az3_spec} and  \ref{fig:Az3_contsub}). In the case of BRI0952, we used the frequency ranges 335.3-336.9, 347.1-347.7, 348.97-348.98, and 350.45-350.67\,GHz. Line emission images were created from the continuum-subtracted data, and continuum images were made from line-free channels. We used Briggs weighting with a robust factor of 0.5 initially for both AzTEC-3 and BRI0925. However, for further analysis of the companion sources surrounding both primary targets, a robust factor of 0.0 was utilized to increase the angular resolution of the \cii emission cubes; these images were subsequently used for the analyses described in this paper. The continuum images were created using a robust factor of 0.5. For BRI0952, the spectral window centered on the \cii emission had a bandwidth of 1.8\,GHz with a spectral resolution of 31\,MHz. The adjacent spectral windows had bandwidths of 2.0\,GHz with spectral resolutions of 15.625\,MHz. For AzTEC-3, the spectral window centered on the \cii emission and the adjacent spectral windows had bandwidths of 2.0\,GHz with a spectral resolution of 31.25\,MHz. Angular resolution and sensitivity is listed in Table~\ref{tab:obs}. 

Polarization image cubes were created at native spectral resolutions of $15.625$~MHz and $31.25$~MHz channels for BRI0952 and AzTEC-3, respectively, using Briggs weighting with a robust factor of $0.5$. For BRI0952, the resulting beam size was $0.51''\times0.40''$ with a position angle of $81^{\circ}$. For AzTEC-3, the mean was $0.91''\times0.64''$ (at $82^\circ$). The different observing conditions and channel width resulted in a channel rms noise in the Stokes Q and U maps of $0.46$~mJy~beam$^{-1}$ and $0.14$~mJy~beam$^{-1}$ for BRI0952 and AzTEC-3, respectively.

For further analysis, including spectral energy distribution (SED) fitting, additional photometry was extracted from archival resources for both AzTEC-3 and BRI0952. \textit{Hubble Space Telescope (HST)} data were extracted from the \textit{HST} archive for both sources. For AzTEC-3, we used the ACS F606W, ACS F814W, WFC3 F105W, WFC3 F125W, and WFC3 F160W bands from the following project codes: 13641 (PI Capak), 9822 (PI COSMOS24-21), and 13384 (PI Riechers). For BRI0952, we used the WFPC2 F814W band with project code 8268 (PI Impey), cross-checked the astrometry with the Gaia2 catalog \citep{Gaia2}, and used photometry from the WISE All-Sky Data Release \citep{Cutri12}. We performed an archive search for the MIPS 24$\mu$m data. We used archival data of ALMA bands 3, 4, and 6 (2017.1.01081.S P.I. Leung and 2015.1.00388.S P.I. Lu). 

\section{Results} \label{sec:results}
\subsection{Polarization} \label{sec:pol}

Our full polarization observations did not reveal any
polarized emission, neither from emission lines nor from the source
continuum. In Table \ref{polres}, we indicate the $3\sigma$ upper
linear polarization ($P_l$) limits that we derive for both sources at
the peak for the \cii line, both for the total integrated line and for the
continuum. 

The linear polarization spectra and continuum values were extracted in a single beam (see Section \ref{sec:pol} for the size) towards the peak of the lines and continuum, respectively. The linear polarization spectrum was produced from the Q and U spectra using $I_p=\sqrt{Q^2+U^2-\sigma_p^2}$, where $\sigma_p=\sqrt{\sigma_Q^2+\sigma_U^2}$ corresponds to the rms error on the linear polarization and $\sigma_{Q,U}$ are the rms errors in Q and U, respectively. Since we detect no polarization, the derived limits include any potential remaining calibration uncertainty. According to ALMA specifications, this $3\sigma$ limit corresponds to $0.1\%$. Thus, the derived upper limits on the \cii lines are
significantly below values that could be expected if the level of
polarization in our sources is similar to that predicted around
Galactic star-forming regions \citep{Zhang2018}.

\begin{figure}[h!]
\centering
\includegraphics[width=8.0cm]{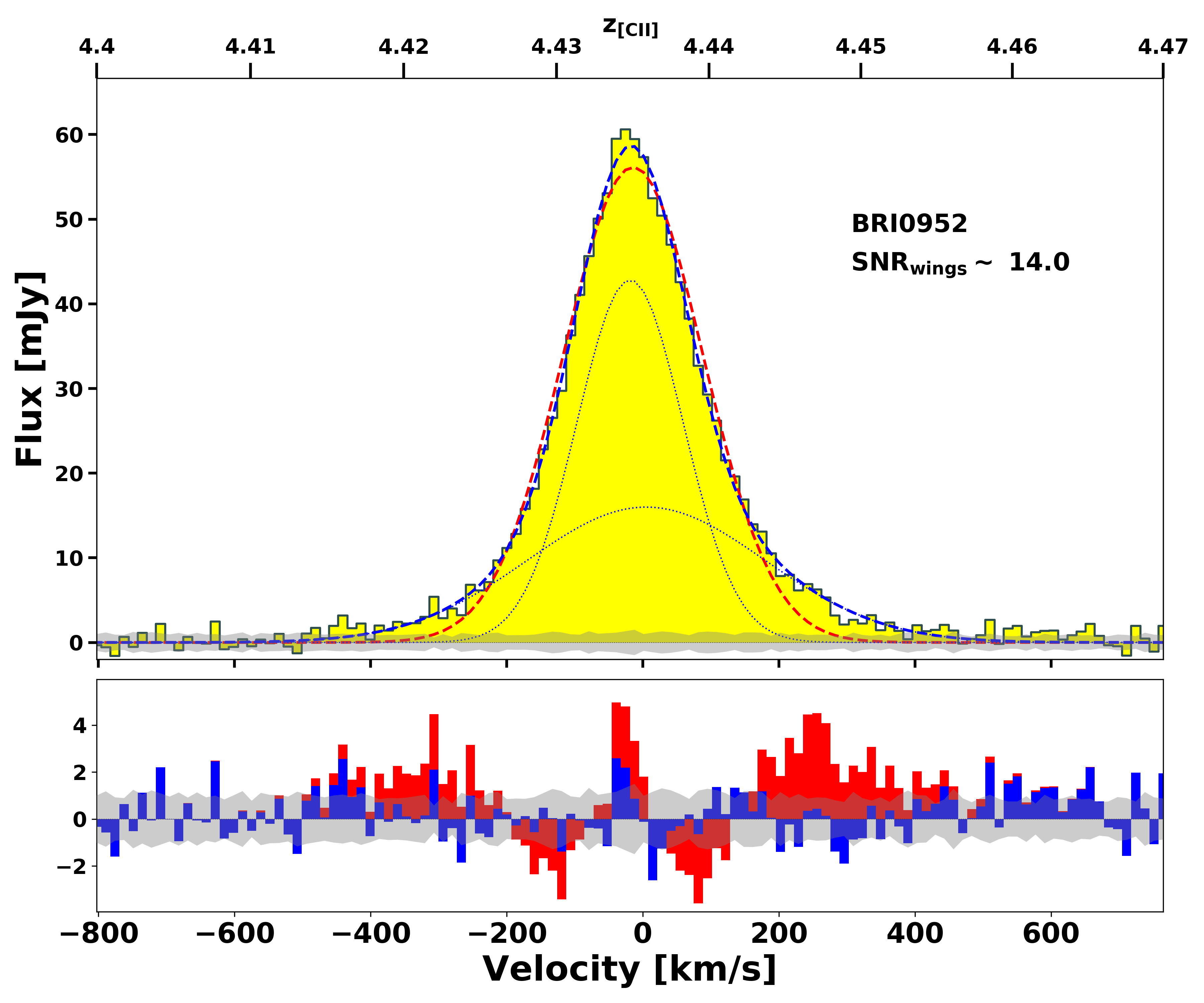}
\includegraphics[width=8.0cm]{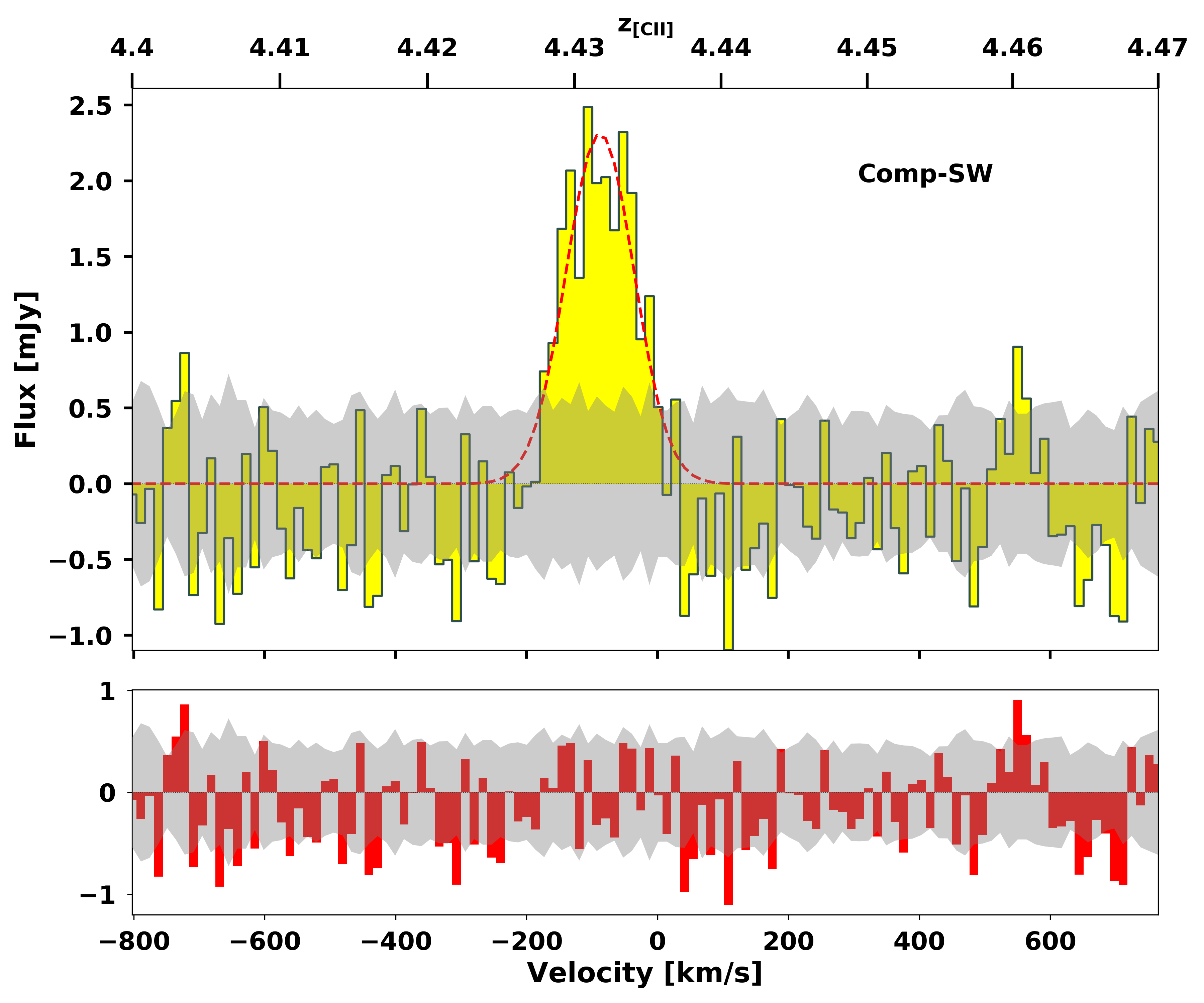}
\includegraphics[width=8.0cm]{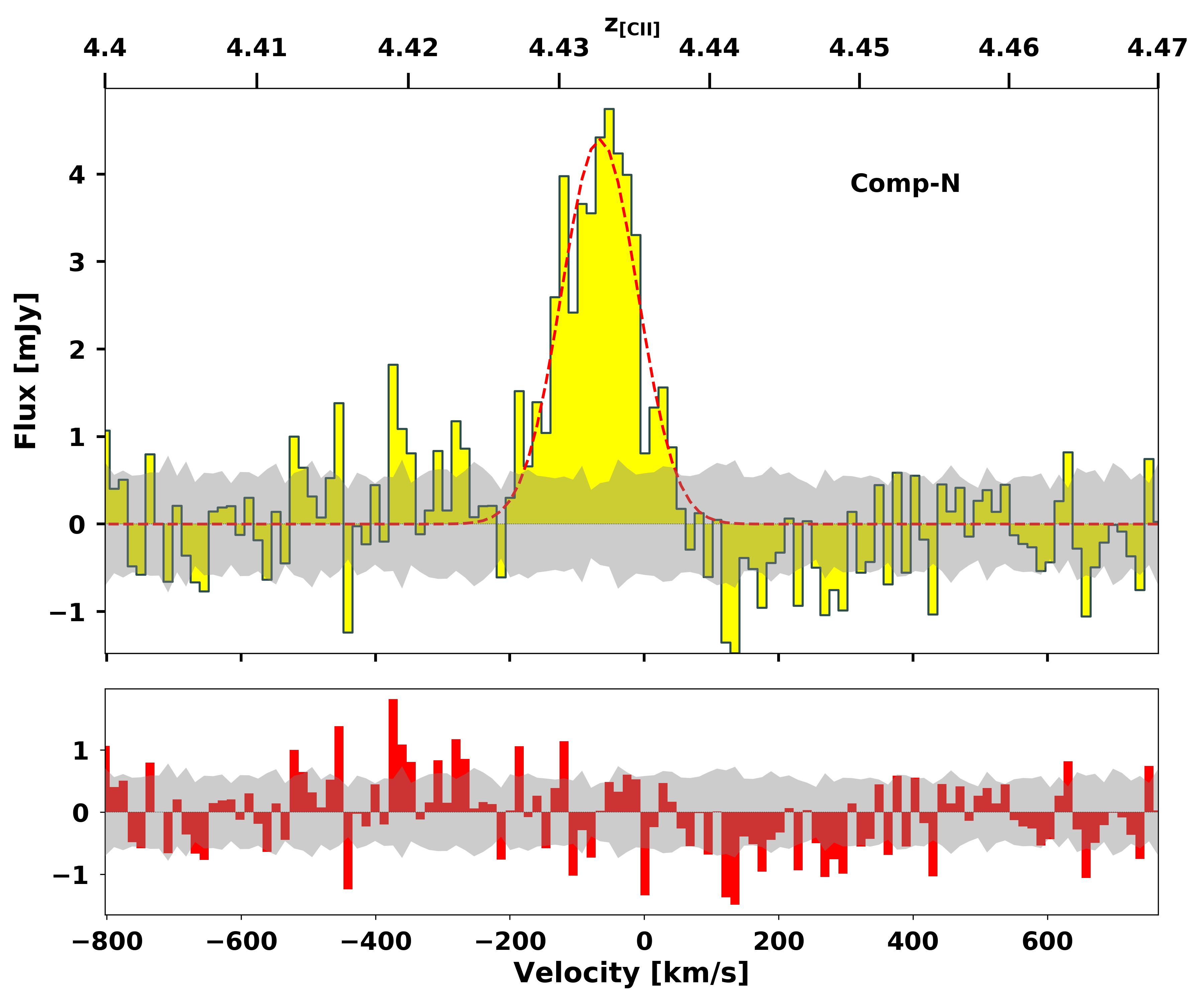}

\caption{Spectra of \cii emission toward BRI0952 and companions. The two companions' line profiles are fit with a single Gaussian, which is shown in red. The quasar's line emission profile is fit with a single Gaussian (red) and a double Gaussian (blue); the double Gaussian corresponding to outflow signatures is clearly seen. The gray region represents the rms of the data in each channel using the procedure described in Section \ref{subsec:BRI_cii}.}.

\label{fig:BRI_spec}
\end{figure} 

\begin{figure*}[h]
    \centering
    \includegraphics[width = 1.0\linewidth]{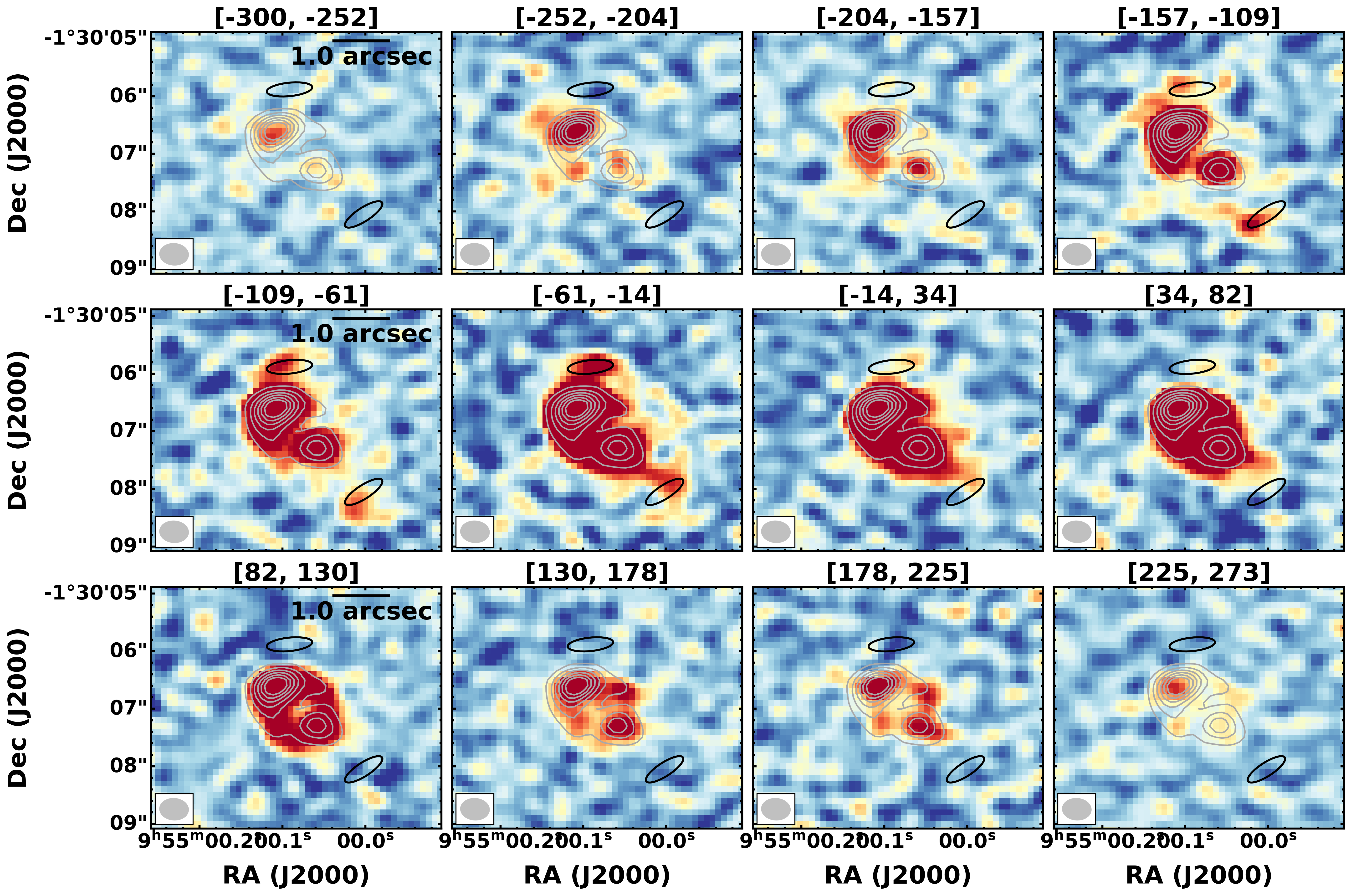}
    \caption{Velocity maps of BRI0952 showing \cii emission across the velocity range of the line. The velocity range is provided at the top of the images. The gray contours are at $10,20,30,40,50,$ and $60\,\sigma$ levels from the moment-0 map of BRI0952, which is shown in Figure \ref{fig:BRI_mom0}. The black ellipses indicate the positions and sizes of the two companions from CASA's {\sc imfit} routine. The southern companion shows a clear velocity gradient across the -157 to -14\,km/s range. The size of the synthesized beam is shown in the bottom left corner of the images.}
    \label{fig:BRI_mosaic}
\end{figure*}

\begin{figure*}[t]
    \centering
    \includegraphics[width = 1.0\linewidth]{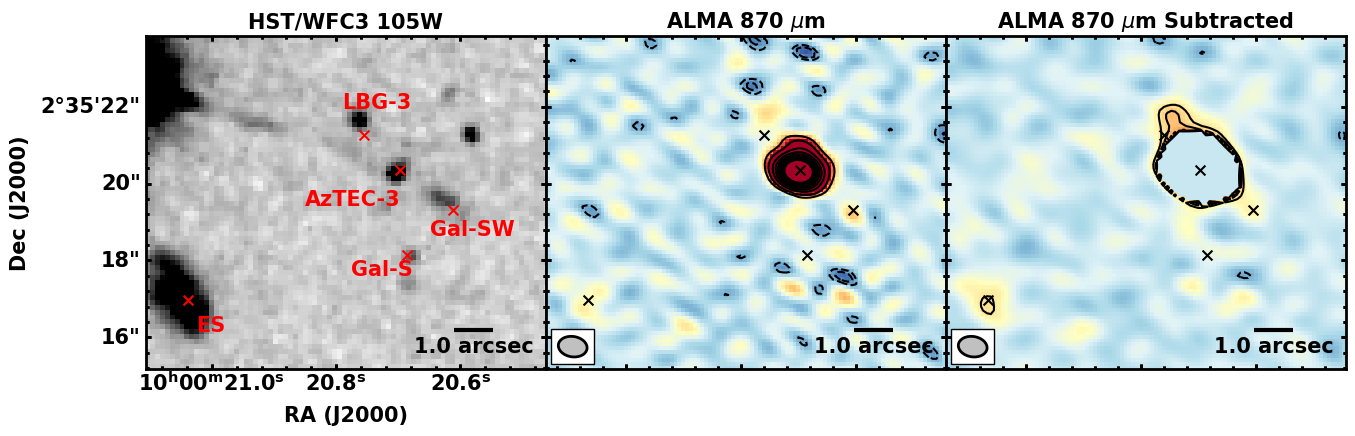}
    \caption{{\it HST}/WFC3 F105W (left) image of AzTEC-3 overlaid with source positions, ALMA $850\,\mu m$ continuum map (center), and SMG subtracted map (right). The continuum maps were created using line-free channels. The residual map was produced following the steps described in Section \ref{subsec:Az3_cont}. The original continuum image contours are shown at $-3,-2,5,10,20,30,40,50,60,70,$ and $80\,\sigma$ levels. The subtracted image contours are shown at $-3,-2,3,4,5,6,$ and $7\,\sigma$ levels. Synthesized beams are shown in the bottom left of the ALMA images.} 
    \label{fig:Az3_info}
\end{figure*}

\begin{figure*}[t]
    \centering
    \includegraphics[width = 1.0\linewidth]{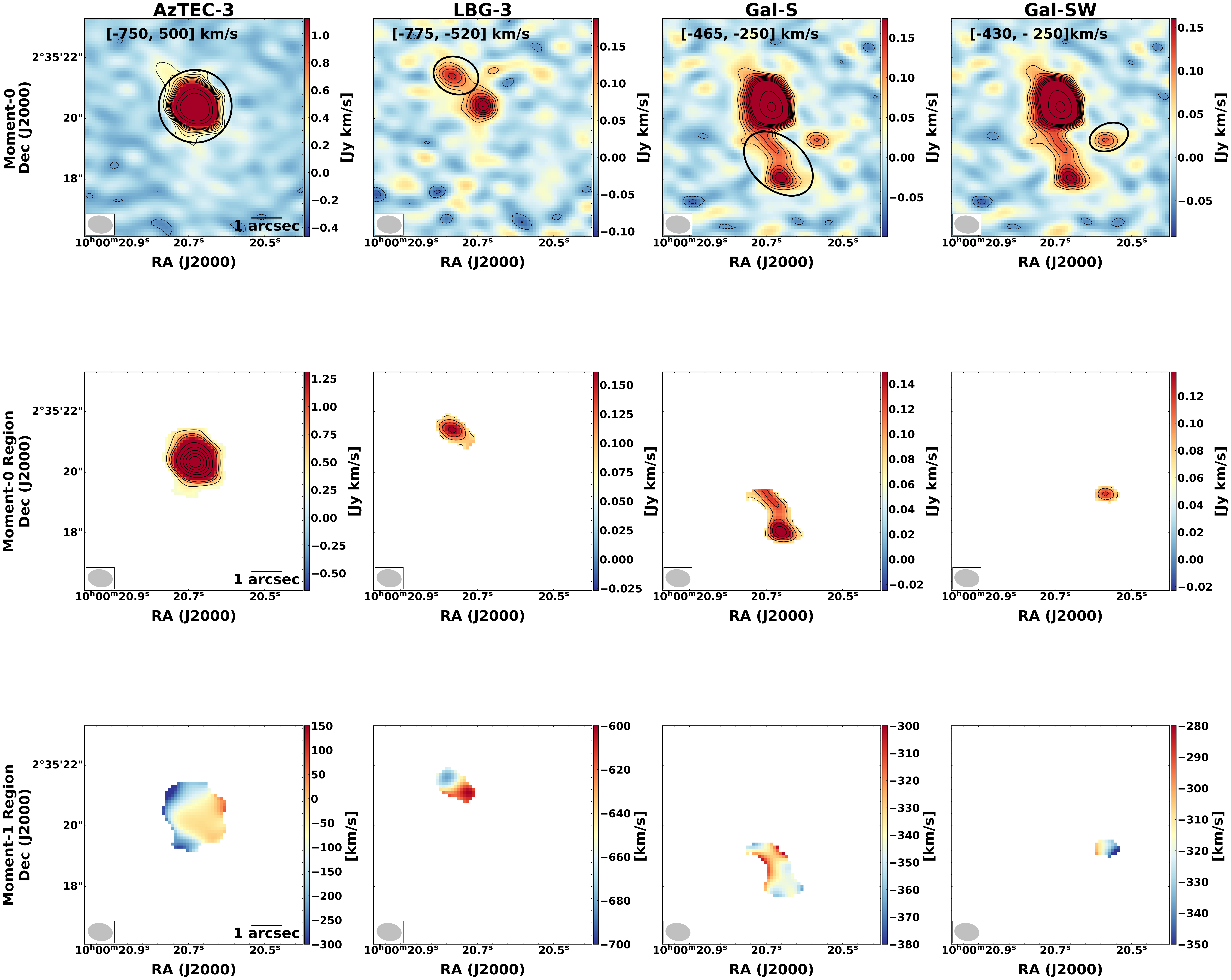}
    \caption{Moment-0 and moment-1 maps for AzTEC-3 system. The top row shows moment-0 maps of the original \cii and the central row shows the moment-0 map of the region from which the spectra were extracted and are centered at the respective redshifts of the sources. The bottom row shows the moment-1 maps from each using the same region that the spectra were extracted from. The contours in the first panel for AzTEC-3 are shown at $-3,-2,10,20,30,40,50,$ and $60\,\sigma$ levels, and the contours shown in subsequent panels are at $-3,-2,3,4,5,6,$ and $7\,\sigma$ levels where the respective $1\sigma$ noise level has been taken from individual moment-0 maps due to the restrictive velocity ranges (and thus differing noise levels). The black circle and ellipses in the top row show the regions from which the spectra were extracted for each source. The synthesized beam is shown in the bottom left of the images.}
    \label{fig:Az3_mom0}
\end{figure*}


\subsection{BRI\,0952-0115} \label{subsec:bri_section}
\subsubsection{Lensing magnification}  \label{subsec:lensing}

The quasar BRI0952 is lensed by a single galaxy, with the lensing source at $z = 0.632$. Although previous lensing models have been constructed for BRI0952 \citep{Lehar00,Eigenbrod07,Gallerani12,Momcheva15}, the wing-like structures we observe in the \cii emission suggested differential lensing across the apparent surface of BRI0952 (seen in Figure \ref{fig:img_model_lensing}), prompting us to create an updated lensing model using our high-resolution data. Additionally, the presence of two companion sources in close proximity to the quasar were also a factor as it was necessary to determine if these represented one double-imaged object or two or more separate sources.  

We determined the lensing parameters of both the source and the lens by utilizing {\sc Visilens}  \citep{Spilker16}. This code is designed specifically to model observations of gravitationally lensed sources at radio and millimeter wavelengths. {\sc Visilens} calculates the magnification factor by directly modeling the \textit{uv} data rather than introducing bias by using images produced from algorithms such as CASA's {\sc CLEAN}. We modeled the lens as a single isothermal ellipsoid parameterized by its location relative to the ALMA image phase center ($x_L$, $y_L$), the ellipticity ($e_L$), and position angle of the major axis ($\phi_L$) in degrees east of north. We modeled the source as a S\'ersic source parameterized by position relative to the lens ($x_S$, $y_S$), flux density, S\'ersic index, half-light radius, axis-ratio, and position angle. These parameters are then run through a Markov chain Monte Carlo (MCMC) fitting procedure, and the best models were output using a deviance information criterion as described in \citet{Spilker16}. 

Initially, we ran {\sc Visilens} on band-7 continuum data in order to determine the best-fit parameters for the lens. Although lens parameters have been previously reported \citep{Lehar00,Eigenbrod07,Gallerani12,Momcheva15}, lens fitting was improved when we allowed the parameters to vary outside of previous values. During the lens optimization, we also allowed the source to vary in position and flux density. Once the lens was optimized, we reran the continuum fit with fixed lens parameters but still allowed the S\'ersic source profile to vary. Similarly, for the band-7 \cii emission data, we ran {\sc Visilens} with the previously optimized lens parameters on the \textit{uv} line emission data while allowing the source to vary in an attempt to increase the quality of our lensing model. The model produced by {\sc Visilens} is a good fit for our data with minimal to no residual emission, and this is shown in Figure \ref{fig:img_model_lensing}. The best-fit parameters are provided in Table \ref{tab:lensing_params}. Furthermore, the magnification factor produced from this model is similar to those previously reported ($\mu \sim {3.92} \pm 1.3$).

The presence of wing-like structures across the \cii frequency range of our BRI0952 observations, as seen in the \cii line emission map (see Figure \ref{fig:img_model_lensing}, panels 1 and 3 for observed \cii data and model output of {\sc Visilens}), prompted consideration of the effect of different lensing factors across the line (i.e., differential magnification). To investigate this, we split the \cii line into five bins constituting five different frequency ranges across the line (these can be seen in Figure \ref{fig:img_model_lensing}) and ran {\sc Visilens}, with the source and lens parameters as specified above, on these bins. The results of this are shown in Figure \ref{fig:img_model_lensing}. We determine that it is likely that some amount of differential lensing across the surface of BRI0952 is occurring, and thus we caution that conclusions drawn that correlate strength of emission with location will be affected by this. However, we conclude that based on the low lensing-magnification factor, this effect will not drastically change the extent or strength of the \cii or continuum emission studied in this paper. 

We also note that images of both the data and residuals, as seen in Figures \ref{fig:img_model_lensing} and \ref{fig:comp_lensing} are produced by {\sc Visilens} rather than by CASA. Although the imaging process in itself is similar to the procedure that CASA performs, it lacks the ability to change weighting schemes and does not clean images. Thus, images produced in {\sc Visilens} are dirty images with natural weighting and therefore will look slightly different than those created by CASA. As a consequence of this, faint sources in close proximity to BRI0952 are not clearly resolved and distinguishable in images produced by {\sc Visilens}. This is in contrast with images produced in {\sc CASA}, which are cleaned with a Briggs weighting scheme using a robust factor of 0.0. The latter are those used throughout the spectral analysis in this paper.

\begin{figure*}[h!]
    \centering
    \includegraphics[width=0.9\columnwidth]{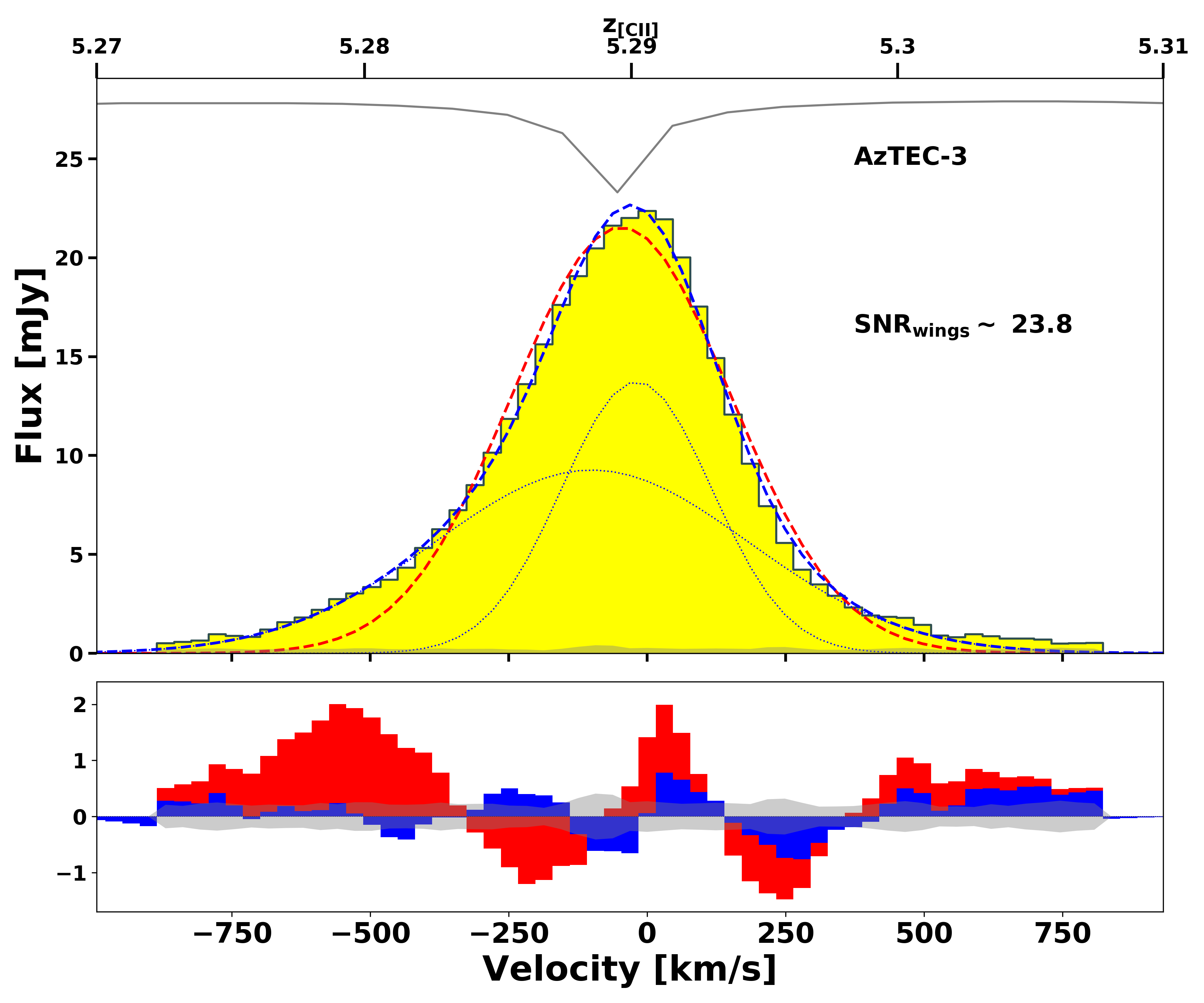}
    \includegraphics[width=0.9\columnwidth]{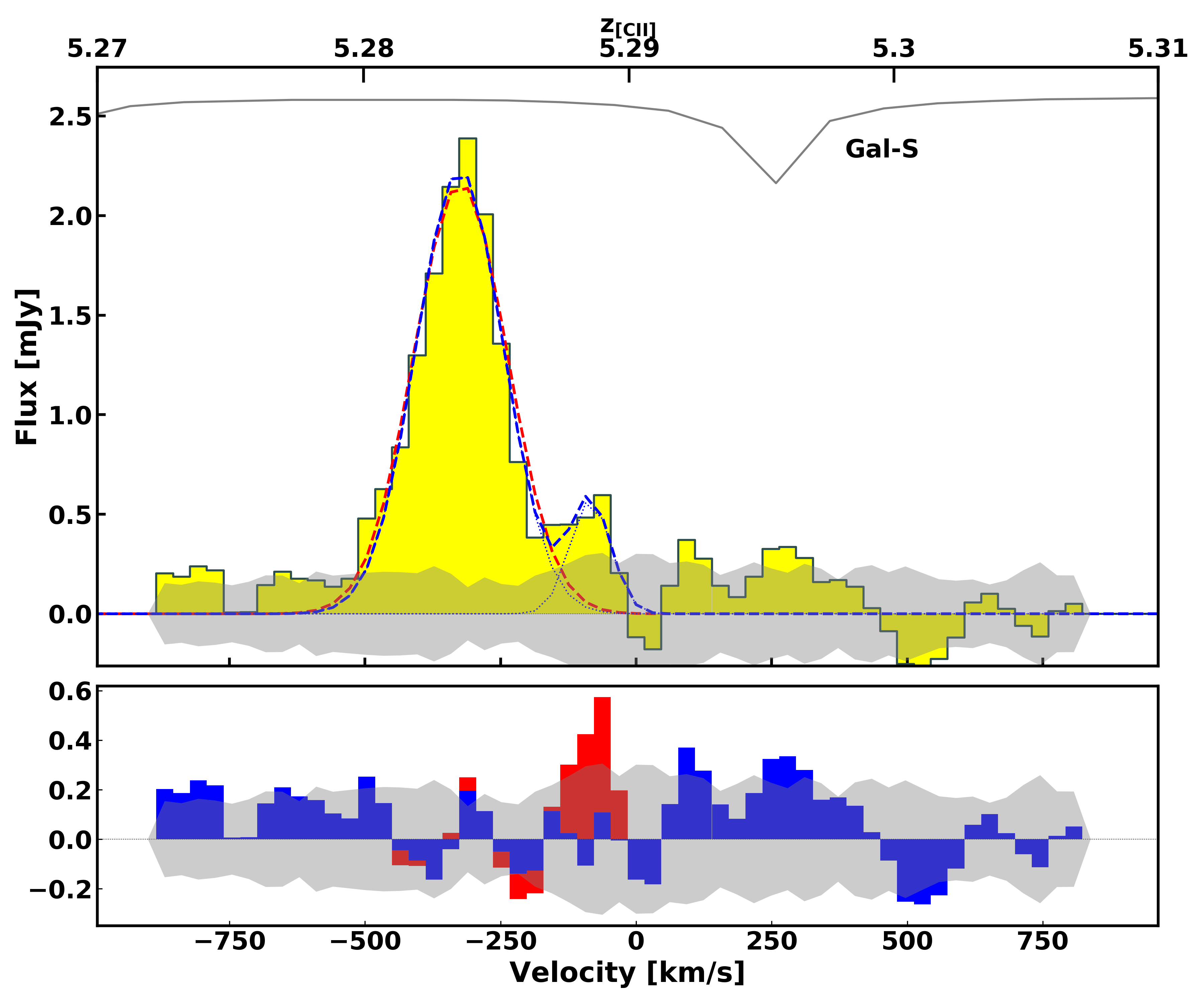}
    \includegraphics[width=0.9\columnwidth]{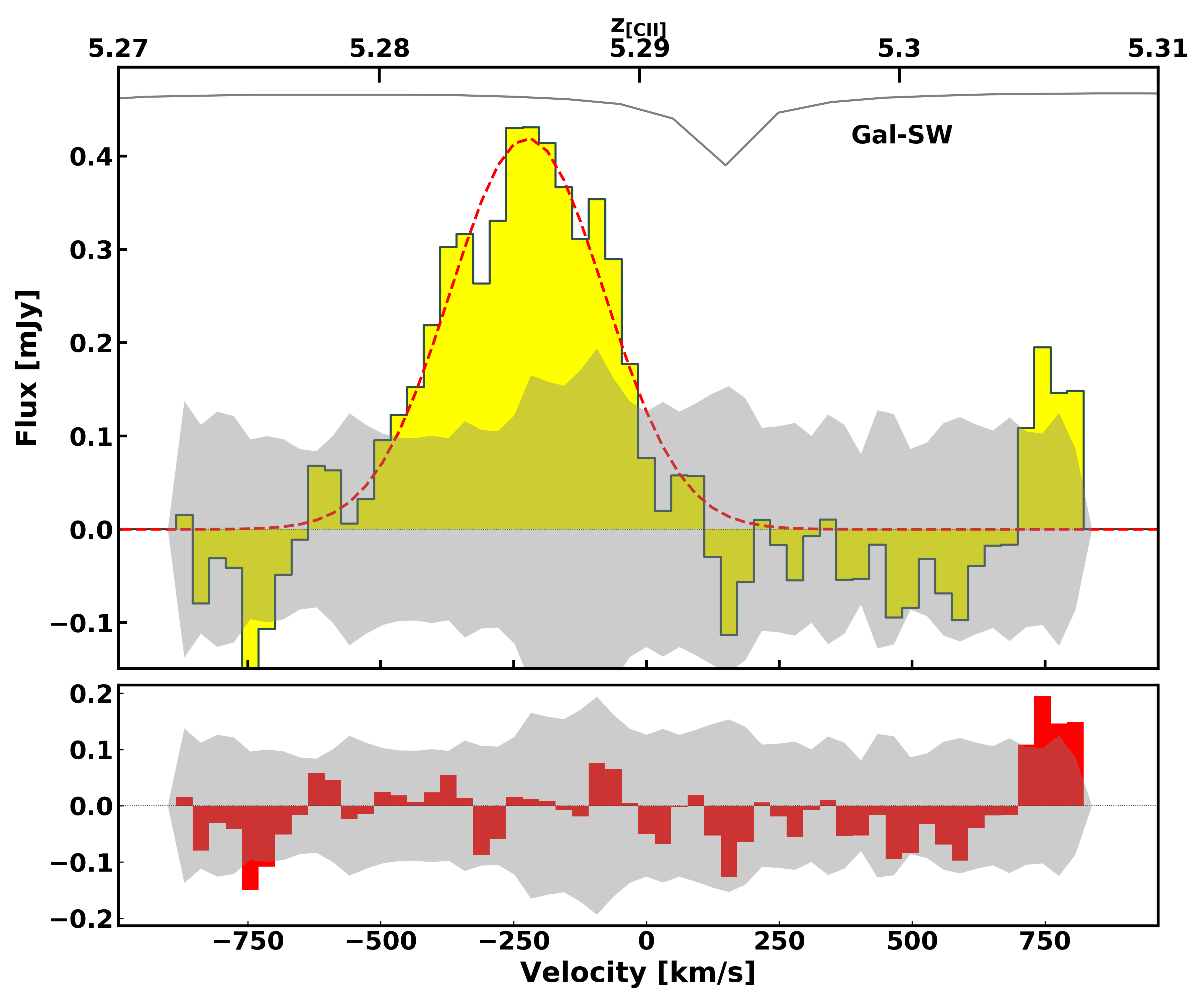}
    \includegraphics[width=0.9\columnwidth]{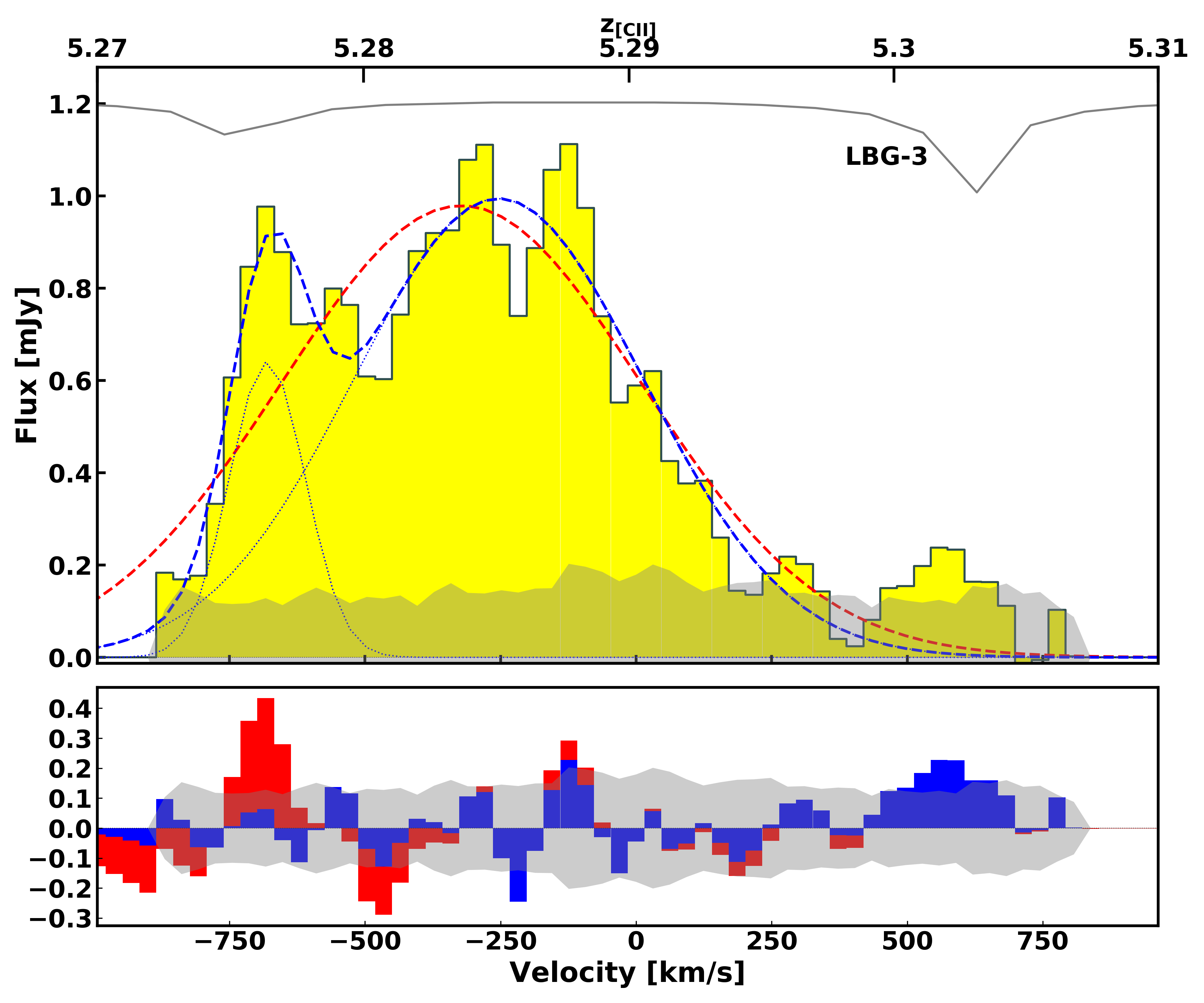}
    \caption{Spectra for the AzTEC-3 system. The red line shows the single Gaussian fit to the data, and the blue line shows the double Gaussian fit to the data. For AzTEC-3, the blue wing $\rm <-500\,km\,s^{-1}$ is significantly more prominent than the red wing. We fit LBG-3 and Gal-SW with double Gaussians to account for the additional "bumps" in the spectra (most obvious in LBG-3), which are likely blended flux from the SMG. The atmospheric transmission is shown as a gray line in the top subplot for every source.}
    \label{fig:Az3_spec}
\end{figure*} 

\subsubsection{\cii emission} \label{subsec:BRI_cii}

We detect \cii (\ce{^{2}P_{3/2}} $\rightarrow$ \ce{^{2}P_{1/2}}) line emission toward both images of the lensed quasar BRI0952. We refer to the two images of the quasar as the "north" and "south" images (names and locations shown in Figure \ref{fig:bri_info}). We confirm the detection of a southern companion source (henceforth termed Comp-SW) tentatively detected by \citet{Gallerani12}; we note that our detection is located in the same region as that of the source reported by \citet{Gallerani12}, but at a significantly lower signal-to-noise ratio (S/N) and a smaller emitting region. Furthermore, we tentatively detect a second companion (termed Comp-N) north of the top image of the quasar. Higher resolution data and a source plane reconstruction of this system would be required to come to a robust conclusion about the nature of this emission. For the purposes of this paper, we treated both sources as additional companion galaxies as the emission is not seen in our lensing model produced by {\sc Visilens,} which suggests it is not simply an affect of the lensing magnification. We isolated our lensing model to only those velocities in which the companions manifest and demonstrate in Figure \ref{fig:comp_lensing} that "residual" emission remains in the regions of the companions, lending further weight to the conclusion that the emission has a separate origin to that of the quasar. Given the relatively lower S/N of Comp-N and Comp-SW in relation to BRI0952, combined with the fact that the lensing model reproduces the flux of BRI0952 using a single S\'ersic brightness distribution, we treated Comp-SW and Comp-N as not sufficiently lensed to affect conclusions drawn in this analysis (in effect $\rm \mu_{comp-N} = \mu_{comp-SW} = 1$).

We note that including additional faint sources in the lensing analysis with {\sc Visilens} did not enable a good fit for the two companion sources; however, given their location, they are likely to be magnified by a factor close to $\mu \sim 1$. Furthermore, based on the close spatial proximity of Comp-SW and Comp-N to BRI0952, we assume the emission is \cii and find spectroscopic redshifts of $z_{\mathrm{Comp-SW}} = 4.4323$ and $z_{\mathrm{Comp-N}} = 4.432$. 

To extract the spectra of BRI0952, Comp-N, and Comp-SW we defined a region around each from which to extract the \cii spectra individually for each component \footnote{This extraction could be done through a number of different methods; however, we find that the outcomes of these methods remain consistent with the method employed in this paper. We also note that methods such as \textit{uv}-plane fitting introduce bias in a similar way to image-plane fitting through base-assumptions of, e.g., a model to describe the source, boundary conditions, and so on.}. For BRI0952, we used an elliptical region of $2.1'' \times 1.45''$ encompassing $\sim 18$ beams\footnote{We define $N_{beams} = A_{extraction}/A_{beam}$.}, containing both Img-S and Img-N of the quasar. For Comp-SW, we used an elliptical region of $1.13'' \times 0.57''$ encompassing $\sim 4$ beams, and for Comp-N we used an elliptical region of $1.3'' \times 0.62''$ encompassing $\sim 5$ beams. These regions are shown in Figure \ref{fig:BRI_mom0} along with moment-0 and moment-1 maps, and extracted spectra are shown in Figure \ref{fig:BRI_spec}. Both companions are faint compared to the emission from BRI0952; both are $\leq 10\%$ of the flux from BRI0952 (see Figure \ref{fig:comb_spec}). The gray plotted rms in Figure \ref{fig:BRI_spec} is calculated in each channel by sampling the spectra in 25 different emission-free regions of the cube extracted from regions the same size as those described above for each individual source.

Due to the presence of high-velocity wings at an $\sim 14 \sigma$ level in the spectra of BRI0952, we find that a double-Gaussian fit was more appropriate (see Figure \ref{fig:BRI_spec}, top panel), yielding an improvement to the reduced $\chi^2$ of more than a factor of four. To investigate the impact of differential magnification on the broad wings and attempt to determine if this feature is simply an artifact from lensing, we calculated the ratio of the flux in the blue and red wings of the line profile. If the ratio of the blue to red wing is more than the error on the lensing factor, we suggest that the broad wings could be due to differential magnification predominantly affecting one side of the line. We define these wings to be between $\pm500$\,km/s and $1/4\times \rm F_{peak}$, where $\rm F_{peak}$ is the peak of the flux. We find the ratio of the red-to-blue wing to be $\sim1.09$, which is less than the error on the magnification factor (1.3). This suggests that while the observed broad velocity wings may be slightly affected by lensing magnification, it is not solely responsible for them. As an additional test, we extracted the spectra from a region only corresponding to Img-N of BRI0952 to determine if the double Gaussian remained a better fit to the data (shown in Figure \ref{fig:BRI_top_spec}). Indeed, the broad wings remain, and a double Gaussian provides an improved fit to the data. We further discuss the implications of this in Section \ref{subsec:outflows}.

We compare the line intensities of each component to those found by \citet{Gallerani12}. We find that our values are lower, both for the intensity of BRI0952 and for Comp-SW (their Comp-C). However, we note a significant discrepancy between their decomposition and ours. \citet{Gallerani12} find that Comp-SW is significantly brighter than either of the individual images of the quasar BRI0952 (here Img-N and Img-S), while we find the opposite to be true. We attribute this to the increased spectral resolution and baseline coverage possible with ALMA over the IRAM Plateau de Bure Interferometer (PdBI). 
 
The observations from \citet{Gallerani12} were carried out with the IRAM Plateau de Bure Interferometer with six antennas in the extended B configuration during three observing runs and the compact C configuration during two observing runs.  They obtained a sensitivity level of 0.5 Jy\,km\,s$^{-1}$\,beam$^{-1}$ in a 300\,km\,s$^{-1}$ channel, which corresponds to a 1$\sigma$ rms of 1.7\,mJy\,beam$^{-1}$. For comparison, our ALMA data for a similar channel width would be $\sim0.1$\,mJy\,beam$^{-1}$, which is about 15-17 times better in terms of sensitivity. We note that the BRI0952 field is near equatorial, meaning that many antennas possibly in combination with several antenna configurations are needed to achieve a good {\it uv}-coverage in the $v$ direction. Given the higher sensitivity combined with the improved {\it uv} coverage thanks to the larger number of ALMA antennas, it is likely that the calibration and image reconstruction of the ALMA data is more robust. As a further comparison between the previous PdBI observations from \citet{Gallerani12} and the new ALMA observations, the spatial resolution is $1.08''\times 0.66''$ and $0.54''\times0.40''$, respectively, and the absolute flux calibration is $\sim20\%$ and $\sim10\%$, respectively. 

We used CASA's {\sc imfit} task to fit the source size for both images of the quasar and companions. We provide \cii line parameters and deconvolved source sizes in Table \ref{tab:CII_line_params}. We note that, due to the lensed nature of BRI0952, this source has a more complex morphology than those typically handled in this way. The source size provided from {\sc imfit} is an indication of the extent of the emission in the image plane and is not necessarily well fit by a Gaussian profile. We added an additional 30\% on the error of the spatial extent of the emission reported for Img-S and Img-N corresponding to the uncertainty on the lensing magnification factor. 

We note the presence of a complex velocity structure between the images of the lensed quasar and the companions. The northern companion exhibits a faint velocity gradient, and the same is present for the south-western companion (see Figure \ref{fig:BRI_mosaic}). The implications of this are further discussed in Section \ref{subsec:environment}. 

\begin{table*}[h]
    \centering
\caption{Model lensing parameters for the foreground galaxy of BRI0952. From left to right: ($x_L$, $y_L$) is the position of the source relative to the ALMA phase center given in the first two columns, with positive values corresponding to west; M is the mass of the lens galaxy; $e_{\rm L}$ is the ellipticity; $\phi_{\rm L}$ is the position of the major axis in degrees east of north; $\gamma$ is the external tidal shear; $\phi_{\gamma}$ is the position angle of the shear, and $\mu$ is the derived lensing factor. Since the values were fixed during fitting of the source position, we do not report their respective errors. The magnification factor is taken to be the average across the line.}
    \label{tab:lensing_params}
    \begin{tabular}{c c c c c c c c c c} \hline \hline
        RA & DEC & $x_L$ & $y_L$ & M & $e_{\rm L}$ & $\phi_{L}$ & $\gamma$ & $\phi_{\gamma}$ & $\mu$ \\
        $\rm [J2000]$ & [J2000] & $\rm ['']$ & $\rm ['']$ & [M$_{\odot}$] &  & (deg. E of N) & & (deg. E of N) & \\ \hline
         09:55:00.1 & -01:30:07.1 & $0.446$ & $0.003$ & $0.619 \times 10^{11}$ & 0.055 & 191 & 0.011 & 63 & $3.92 \pm 1.3$ \\ \hline
    \end{tabular}
\end{table*}

\subsubsection{Continuum} \label{subsec:bri_cont}

We detect strong continuum emission toward both north and south images of BRI0952. Using CASA's {\sc imfit,} we fit both the north and south images of the quasar and report strong continuum emission totaling $1.96 \pm 0.43$ mJy, which is corrected for lensing. Continuum fluxes from individual images of the quasar are provided in Table \ref{tab:cont} and images are shown in Figure \ref{fig:bri_info}. We do not detect the companion sources in continuum and report a $3\sigma$ upper limit of $0.13$ mJy for each. Due to the non-detection of the companions in the continuum, we defined a region corresponding to the quasar from which to extract the continuum flux. As for the line emission, the source sizes of BRI0952 reported in Table \ref{tab:cont} are reported with an additional 30\% error added to account for the complex morphology and lensing uncertainties. 

Since we do not detect the companions in continuum emission, we estimated their flux using relations from their $L_{\rm{[C\,{\sc II}]}}$ to determine if they should be detectable in our ALMA observations. We estimated their SFR's using the relation $\rm log(SFR)\,[M_{\odot}/yr] = 1.0 -7.06\times log(L_{[CII]})\,[L_{\odot}]$ from \citet{DeLooze14} for starburst galaxies, computed their $L_{\rm{IR}}$ using the relation in Section \ref{subsec:sfr} \citep{Carilli13}, and used a modified blackbody approximation \citep[e.g.,][Equation 2]{Knudsen03} to recover their $S_{350 \rm{GHz}}$. This results in $S_{350 \rm{GHz}} = 0.21$ mJy for the northern companion and $S_{350 \rm{GHz}} = 0.10$ mJy for the south-western companion. This would correspond to an $\sim4\sigma$ detection for the northern companion and an $\sim 2\sigma$ detection for the southern companion. Thus, in the case of the southern companion with the current data a continuum emission detection is unfeasible. This may also be the case for the northern companion. However, it is also possible that without the velocity information encoded in emission line spectra (and thus without the ability to severely isolate the frequency or velocity range from which to attempt to extract information about the emission from the companions), our observations are simply insufficient to resolve any continuum emission from the companions. 

\subsubsection{SED fitting} \label{subsec:bri_SED}

We performed IR SED fitting and decomposition of the AGN and star formation contributions, using DecompIR \citep{Mullaney11} following the methods of \citep{Stanley18}. We used photometry from an archival search for WISE bands and MIPS 24\,$\mu m$. We also used ALMA archival data from bands 3, 4, and 6. We performed two sets of fits, one with only the star formation templates and one with both AGN and star formation templates. We find that the photometry of BRI0952 is best fit by a combination of AGN and star formation emission in the IR. The decomposition of the IR SED allows for a calculation of the SFR without contamination from the AGN emission, using the $L_{\rm IR_{SF}}$ integrated from 8-1000 $\mu$m. IR luminosities are reported in Table \ref{tab:derived_props} and discussed further in Section \ref{subsec:LIR}. 

\subsection{Aztec-3} \label{subsec:aztec3}

\subsubsection{\cii emission} \label{subsec:Az3_cii}

We detect \cii (\ce{^{2}P_{3/2}} $\rightarrow$ \ce{^{2}P_{1/2}}) emission toward the SMG AzTEC-3. We report three sources (LBG-3, Gal-S, and Gal-SW) of the same emission in the region shown in Figure \ref{fig:Az3_info}. LBG-3 was initially detected as an Lyman-break galaxy with the COSMOS survey \citep[optical ID 1447526]{Ilbert09} but was undetected in \cii in \citet{Riechers14} and the authors presented a $3\sigma$ upper limit on its \cii emission. An additional companion, LBG-2 - not shown in Figure \ref{fig:Az3_info}, was also originally detected by COSMOS as an LBG but was also undetected in \cii by \citet{Riechers14}. Similarly, we do not detect the source in \cii emission, and thus do not investigate this source further. Gal-S and Gal-SW are first detections. 

The data of \citet{Riechers14} were obtained during Cycle-0 with 16-24 antennas, with a total observing near 7000\,seconds spread evenly over two pointings. As a result, the sensitivity of the data in the region around AzTEC-3 is lower by about a factor of two compared to the new data presented here; however, it is important to note that the beam area of our data is 1.7$\times$ larger.  We note that the $3\sigma$ limit from \citet{Riechers14} appears to be for a single beam, while we obtain a detection over a more extended region corresponding to $2.8\times$ the size of beam used in this analysis.  
Furthermore, we note that if converting their $3\sigma$ upper limit to a $5\sigma$ upper limit, the resulting $I_{\rm [CII]}$ is consistent with our detection, though it is important to keep in mind that in this comparison we do not take into account the integration over an extended area.  

We assume that the emission from the companions is \cii based on their close proximity to AzTEC-3, and find a spectroscopic redshift of $z_{\rm LBG-3} = 5.284$, which is in relative agreement with the previous spectroscopic redshift of $z = 5.3$ \citep{Riechers14}. We report spectroscopic redshifts for the additional companions, $z_{\rm Gal-S} = 5.2919$ and $z_{\rm Gal-SW} = 5.2942,$ respectively. Figure \ref{fig:comb_spec} shows the combined spectra of the companions and AzTEC-3.

\begin{table}[h]
    \centering
    \caption{Continuum fluxes for both systems. The continuum fluxes are provided for both images of BRI0952 and are corrected for lensing. However, we do not correct the sizes for lensing. We note that the error on the companions of AzTEC-3 includes an additional 10\% due to the additional uncertainty introduced from the deblending process and the errors on the size of both images of BRI0952 include an additional 30\% to account for lensing uncertainty.}
    \label{tab:cont}
    \begin{tabular}{l c c } \hline \hline
        Galaxy & Region & $S_{\nu}$ \\
         & ($'' \times ''$) & [mJy] \\ \hline 
        $\rm BRI0952_{N}$ & ([$0.35\pm0.15] \times [0.20\pm0.13$]) & $1.44\pm0.4$ \\
        $\rm BRI0952_{S}$ & ([$0.31\pm0.15] \times [0.26\pm0.13$]) & $0.52\pm0.14$ \\ 
        Comp-N & - & $<0.13$ \\
        Comp-SW & - & $<0.13$ \\ \hline
        AzTEC-3 & ([$0.38\pm0.03] \times [0.27\pm0.07$]) & $6.06 \pm 0.13$ \\
        LBG-3 & ([$0.59\pm0.07] \times [0.35\pm0.07$]) & $0.19 \pm 0.02$ \\
        Gal-SW & ($<0.91 \times <0.66$) & $0.07 \pm 0.02$ \\
        Gal-S & - & $<0.056$ \\
        ES & ([$1.09\pm0.38] \times [0.37\pm0.36$]) & $0.21 \pm 0.05$ \\ \hline
    \end{tabular} 
\end{table}

We followed the same procedure as described for BRI0952 and individually extracted the spectra for the sources in this system from a region with a very limited velocity range \footnote{Again, similar to BRI0952; if other methods are used, the results remain consistent with the regional spectral extraction used here.}. We used a circular region with a radius of $2.4''$ to extract the spectra of AzTEC-3 encompassing $\sim 13$ beams. We used elliptical regions for the following companions: $1.2'' \times 1.5''$ for LBG-3 encompassing $\sim 4$ beams, $2.6'' \times 1.7''$ for Gal-S encompassing $\sim 10$ beams, and $1.3'' \times 0.90''$ for Gal-SW encompassing $\sim 3$ beams. These regions are shown in Figure \ref{fig:Az3_mom0}. We were able to minimize the source blending of the companions with AzTEC-3 since the spectra of the companions are extracted over a very limited velocity range, as shown in Figure \ref{fig:Az3_mom0}. We calculated the rms in the same way as described in Section \ref{subsec:BRI_cii}. Additionally, we include the atmospheric transmission in the top subplot for every source in Figure \ref{fig:Az3_spec} to show where the atmospheric absorption line is (shown in Figure \ref{fig:Az3_contsub}), specifically in relation to the companion spectra.

\begin{figure*}
    \centering
    {\includegraphics[width = 1.0\linewidth]{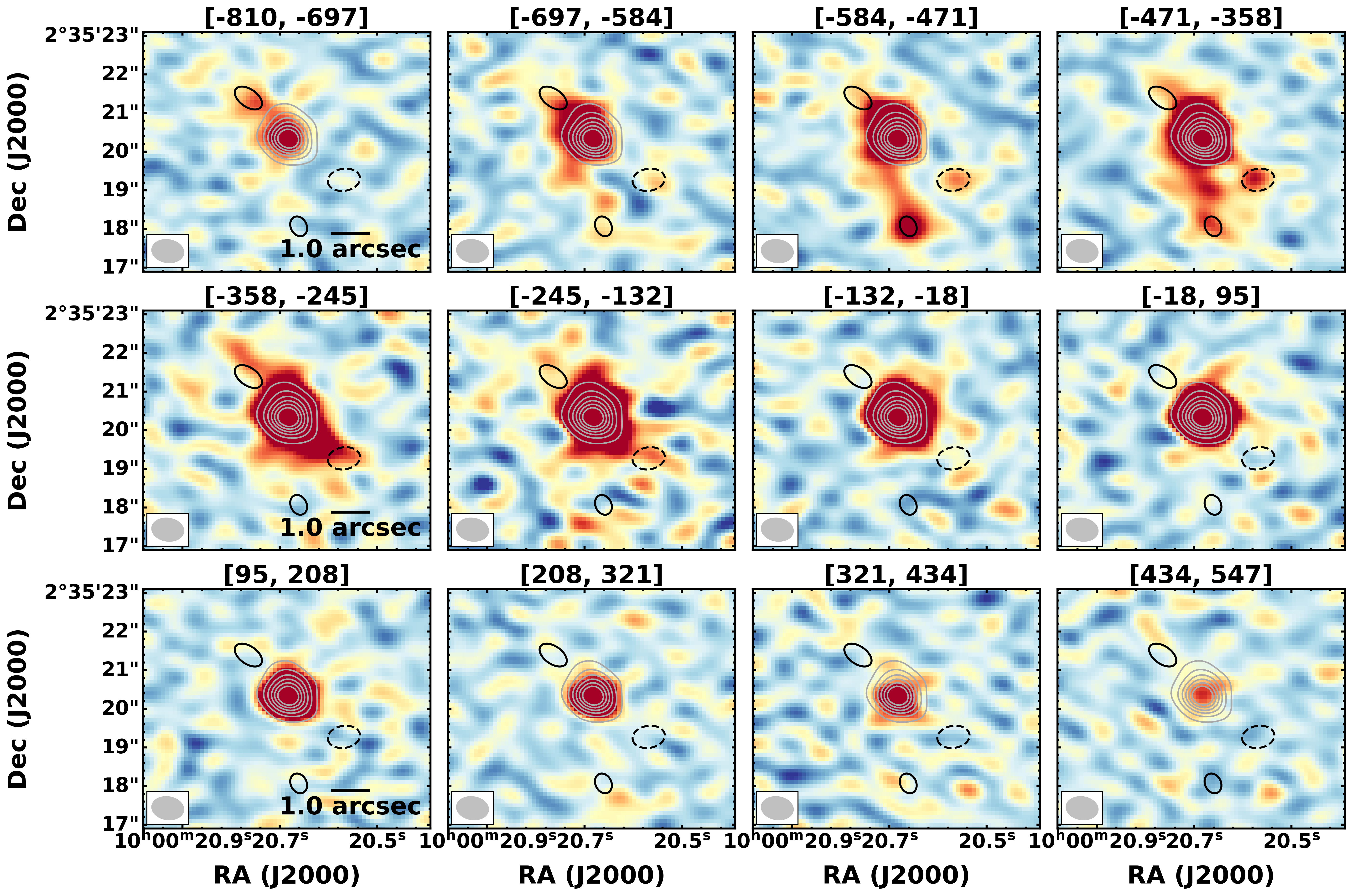}}
    \caption{Velocity maps of AzTEC-3 showing \cii emission across the velocity range of the line. The velocity range is provided at the top of the images. The gray contours are at $10,20,30,40,50,$ and $60\,\sigma$ levels from the moment-0 map of AzTEC-3, which is shown in Figure \ref{fig:Az3_mom0}. The black ellipses indicate the positions and sizes of the three companions from CASA's {\sc imfit} routine where the dashed ellipse around Gal-SW denotes an upper-limit point source. The "gas-bridge" structure between AzTEC-3 and Gal-S is seen in the velocity range of -584 to - 358\,km\,s$^{-1}$. The size of the synthesized beam is shown in the bottom left corner of the images. }
    \label{fig:Az3_mosaic}
\end{figure*}

The \cii spectral line profile of AzTEC-3 shows the presence of high-velocity wings at a $\sim 24 \sigma$ level -and the profile is better fit with a double Gaussian to account for this (Figure \ref{fig:Az3_spec})- where the reduced $\chi^2$ is improved by more than a factor of four using a double-Gaussian fit. We note that the blue wing is wider compared to that of the red wing, as reported in \citet{Riechers14}. We used CASA's {\sc imfit} to determine deconvolved source sizes. We report source sizes and Gaussian line parameters for AzTEC-3 and companions in Table \ref{tab:CII_line_params}. 
The \cii emission from the central source is emitted over a compact region with an extent of $4.2 \pm 0.4$\,kpc along the major axis, which is consistent with the emitting region of $3.9$\,kpc reported by \citet{Riechers14}. We do not detect a strong velocity gradient, which is also in good agreement with \citet{Riechers14}. 

We note an additional feature: a possible velocity gradient between AzTEC-3 and LBG-3. We show this in Figure \ref{fig:Az3_mosaic} and Figure \ref{fig:lbg3_velgrad}. This is a tentative feature as it only appears in a few spectral elements and could simply be emission associated with AzTEC-3. \citet{Guaita22} detected a bridge of Lyman-$\alpha$ emission extending between these two components, indicating that tidal forces could be at work between LBG-3 and AzTEC-3. The detection of Lyman-$\alpha$ in both AzTEC-3 and LBG-3 further supports the likelihood of an interacting system. This is further discussed in Section \ref{subsec:environment}.

\begin{table*}
    \small
    \centering
    \caption{Line parameters for the \cii emission lines. The distance is measured between the central source (AzTEC-3 and the northern image of BRI0952, respectively) and the surrounding sources. $I_{\rm [CII]}$ is the line intensity, and FWHM is the full width at half maximum of the lines (one entry for single-Gaussian fit and two entries for double-Gaussian fit). $A_{\rm [CII]}$ is the length of the major axis of the deconvolved source size from CASA's {\sc imfit}, we note that BRI0952 has a more complex morphology, and thus the sizes include an additional 30\% error on the reported value corresponding to the uncertainty on the lensing factor.}
    \label{tab:CII_line_params}
    \begin{tabular}{l c c c c c c c} \hline \hline
        Name & R.A. & Dec. & Distance & $z_{\rm [CII]}$ & $I_{\rm [CII]}$ & $\rm FWHM_{[CII]}$ & $A_{\rm [CII]}$ \\ 
         & [J2000] & [J2000] & [$''$] & & [Jy km $\rm s^{-1}$] & [km $\rm s^{-1}$] & [$''\times ''$] \\
        \hline 
        
        BRI0952 & N: 09:55:00.10 & $-$01:30:06.61 & & $4.433$ & $15.45 \pm 1.5$ &  $ 182 \pm 5$ & ([$0.51 \pm 0.19] \times [(0.285 \pm 0.12$]) \\
         & S: 09:55:00.06 & $-$01:30:07.29 & 1.0 & & &  $410 \pm 21$ & ([$0.36 \pm 0.13] \times [0.29 \pm 0.21] $])\\
        
        Comp-N & 09:55:00.09 & $-$01:30:05.89 & 0.75 & $4.432$ & $0.62 \pm 0.06$ & $130 \pm 10$ & ([$0.79 \pm 0.32] \times [0.23 \pm 0.11$]) \\
        
        Comp-SW & 09:55:00.00 & $-$01:30:08.05 & 2.2 & $4.432$ & $0.30 \pm 0.04$ & $122 \pm 13$ & ([$0.76 \pm 0.30] \times [0.23 \pm 0.13] $]) \\ \hline
        AzTEC-3 & 10:00:20.696 & +02:35:20.35 & & $5.2988$ & $ 11.34 \pm 1.0 $ & $ 320 \pm 12 $ & ([$0.69 \pm 0.06] \times [0.42 \pm 0.10 $]) \\
        & & & & & & $660 \pm 20$ & \\
        
        LBG-3 & 10:00:20.766 & +02:35:21.39 & 1.5 & $5.2841$ & $0.78 \pm 0.07$ & $630 \pm 36$ &  ([$0.8 \pm 0.02] \times [0.46 \pm 0.02$]) \\
        & & & & & & $165 \pm 24$ & \\
        
        Gal-S & 10:00:20.68 & +02:35:18.07 & 2.3 & $5.2919$ & $0.52 \pm 0.05$ & $190 \pm 10$ & ([$0.53 \pm 0.10] \times [0.41 \pm 0.15 $]) \\
        & & & & & & $92 \pm 23$ & \\
        
        Gal-SW & 10:00:20.60 & +02:35:19.28 & 1.8 & $5.2942$ & $0.15\pm0.01$ & $325 \pm 24$ & ([$0.36 \pm 0.04] \times [0.08 \pm 0.10$]) \\
        \hline
    \end{tabular}
\end{table*}
\subsubsection{Continuum} \label{subsec:Az3_cont}
We obtain continuum emission from imaging line-free channels in the band-7 ALMA data. Due to the lower angular resolution combined with the lack of velocity information provided in the spectral emission, we were unable to isolate the emission from the companions by restricting the velocity range from which they were extracted as we did for the \cii emission. Hence, we extracted the continuum flux using three different methods.

The first was to use CASA's {\sc imfit} routine on the marginally resolved continuum image. From this we obtain a continuum flux of $6.06 \pm 0.13$\,mJy for AzTEC-3, which is in good agreement with \citet{Riechers14}, $0.071 \pm 0.018$\,mJy for Gal-SW, and $0.18 \pm 0.023$\,mJy for LBG-3.

Secondly, we used the {\sc imfit} residual image with the continuum emission from AzTEC-3 removed to obtain the continuum flux of the companions. Using this method, we obtain a continuum flux of $0.082 \pm 0.02$\,mJy for Gal-SW and $0.19 \pm 0.024$\,mJy for LBG-3. 

Finally, we attempted to isolate the origin of the continuum flux for the companions by creating a continuum image corresponding only to the flux from AzTEC-3, which we were then able to subtract from the original continuum image. To model the flux from AzTEC-3, we selected emission that was at levels higher than $9\sigma$. This emission was then subtracted from the continuum image. The original continuum image and the subtracted image are shown in Figure \ref{fig:Az3_info}. This resulted in a continuum flux of $0.06 \pm 0.017$\,mJy for Gal-SW and $0.22 \pm 0.025$\,mJy for LBG-3. Due to the similarity of the results from various methods, we conclude that the intrinsic value of the continuum for each is in the range provided for the continuum for LBG-3 and Gal-SW.

We do not detect Gal-S in continuum, and we report a $3 \sigma$ upper limit of $0.05$\,mJy. The companion sources are faint compared to AzTEC-3 ($\sim3\%$ for LBG-3 and $\sim1\%$ for Gal-SW), and their flux contribution to the central source is negligible. Flux measurements and deconvolved source sizes for the field are provided in Table \ref{tab:cont}, where average values are provided for LBG-3 and Gal-SW for both continuum flux and size estimates.

We note an offset between the ALMA continuum, \textit{HST}, and \cii emission for LBG-3. This is shown in Figure \ref{fig:cont_offset}. We suggest that this offset between the three types of emission is likely linked to spatial offset between different types of emission in this source. The offset of this emission also affects the subtraction we performed above due to the possibility that we have in fact subtracted some flux pertaining to LBG-3 that could be included in our model of the emission from AzTEC-3. Thus, we suggest that the continuum flux reported in this paper for LBG-3 be treated as a lower limit. Without additional higher resolution data, a robust conclusion about the nature of this offset is not possible.

We detect an additional continuum source at $\sim10.5$ significance, referred to extra source (ES), in the AzTEC-3 field at the position 10:00:21.066, +02:35:16.975, which is shown in Figure \ref{fig:Az3_info}. This source is bright across the {\it HST} WFC3 filters, but it is undetected in \cii emission, indicating that it is at a different redshift. Continuum flux measurements and source size are provided in Table \ref{tab:cont}.

\subsubsection{SED Fitting} \label{subsec:Az3_SED}
We fit the SED of AzTEC-3 using photometry extracted from \textit{HST} imaging utilizing the source extraction algorithm SExtractor \citep{Sextractor} and our ALMA photometry (see Table \ref{tab:az3_phot}). We matched the resolution of the \textit{HST} images to the lowest resolution filter; we note that the companion sources are not blended with the emission from AzTEC-3 in the \textit{HST} imaging. We fit the SED using Bagpipes \citep{Carnall18}, assuming a constant star formation history, a Calzetti reddening law to describe the dust attenuation, and a Chabrier initial mass function \citep[IMF,][]{Chabrier03}. The mass of the central source was allowed to vary from $10^8 - 10^{14} \rm M_{\odot,}$ and the metallicity was allowed
to vary from 0.01-1.0 $\rm Z_{\odot}$. We used the SED fit from Bagpipes to calculate the $L_{\rm IR}$ (integrated from 8-1000 $\mu$m) of AzTEC-3, which is subsequently utilized below in determination of the SMG's properties. 

\begin{figure*}[h]
\centering
\begin{minipage}{8cm}
{\includegraphics[width=8 cm, height = 5cm]{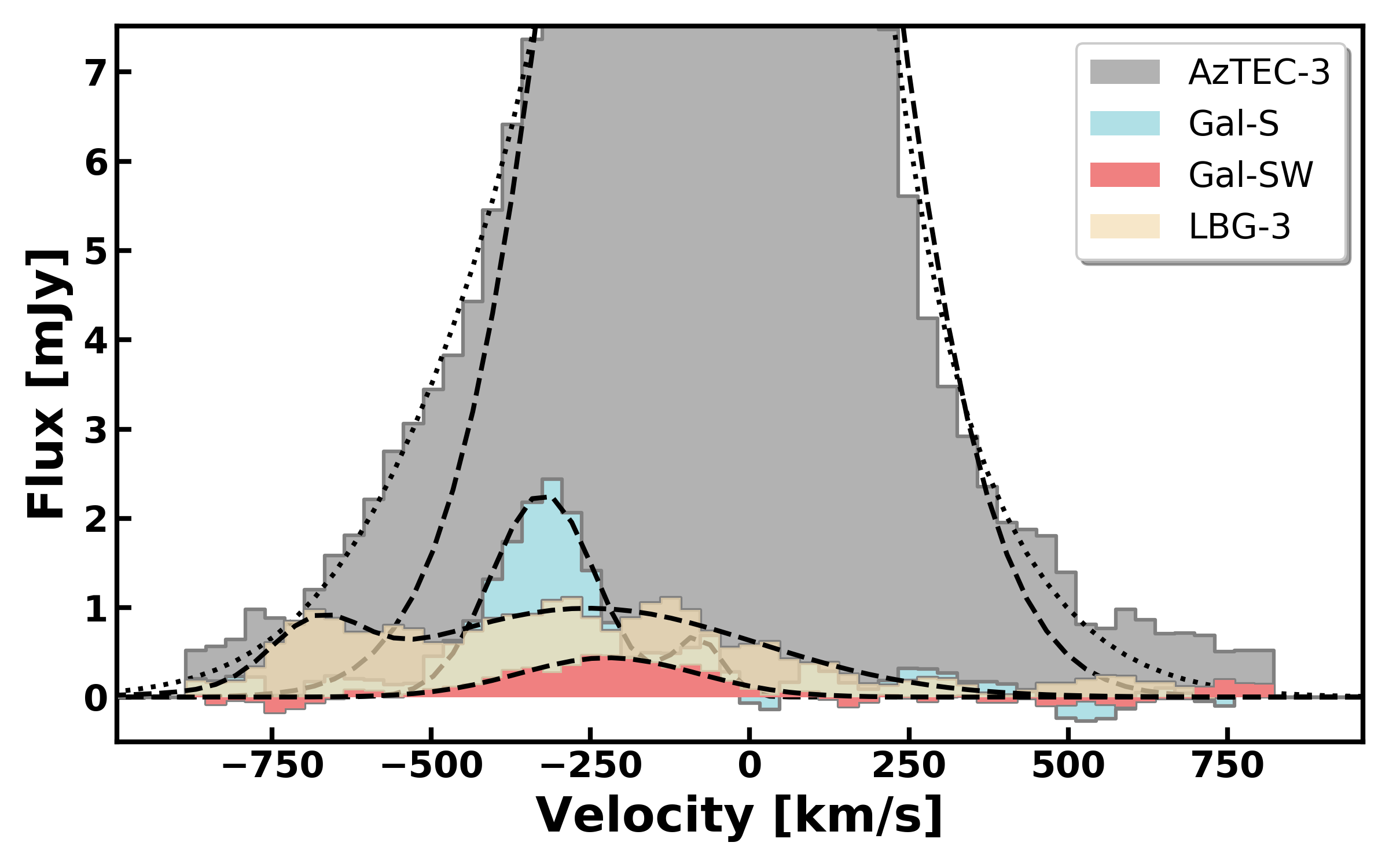}}
\end{minipage}
\begin{minipage}{0.01cm} \mbox{ }
\end{minipage}
\begin{minipage}{8cm}
\includegraphics[width=8 cm, height = 5cm]{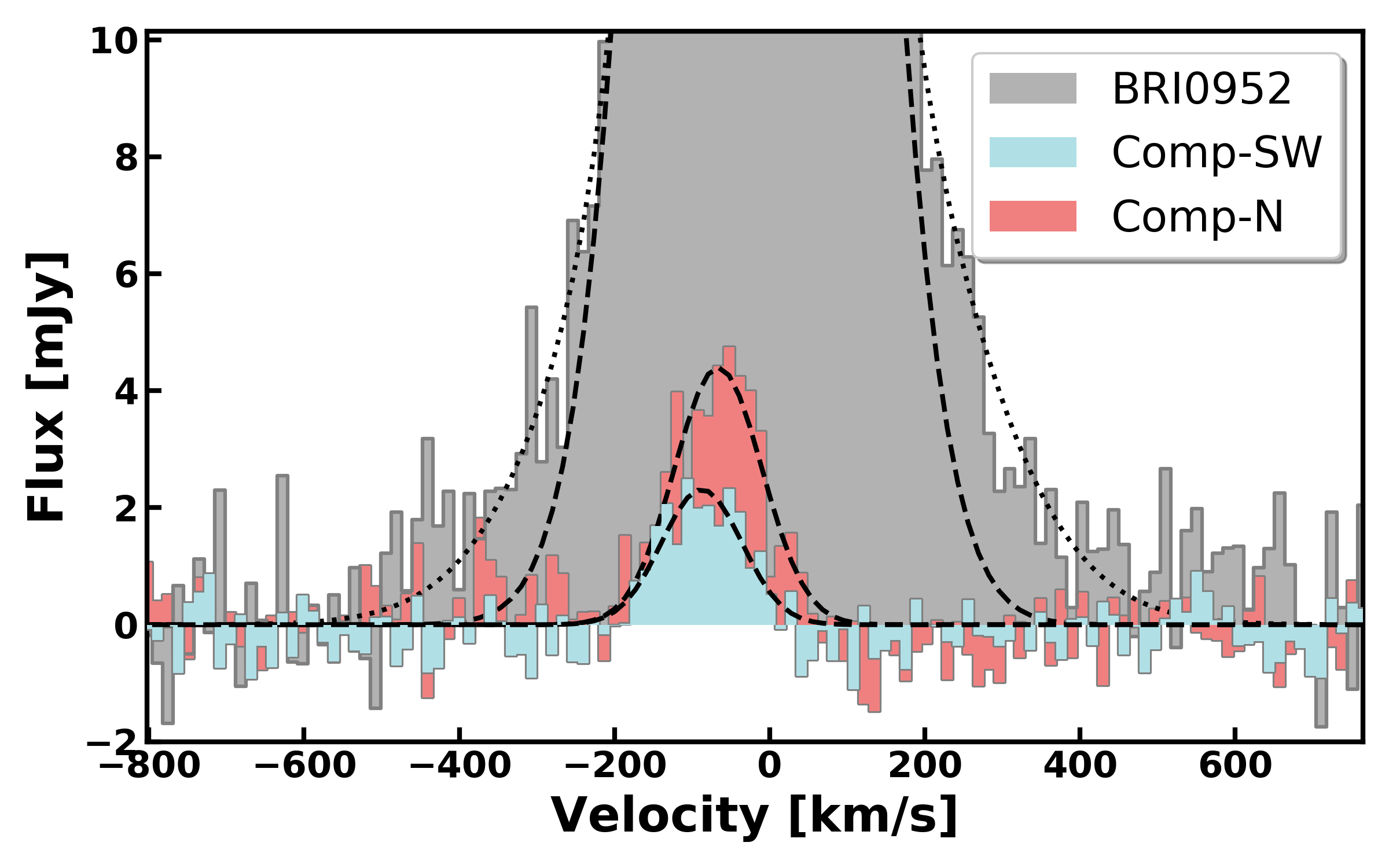}
\end{minipage}
\caption{\cii spectra for the companions of BRI0952 and AzTEC-3. The spectra of both the quasar and the SMG have been truncated so that the companion spectra are easily seen. The single- (dashed) and double- (dotted) Gaussian fits for both AzTEC-3 and BRI0952 are overplotted, along with the Gaussian fits to the companion sources. We highlight that the companions to BRI0952 are located at a very similar systemic velocity, whereas AzTEC-3's companions are located in the blue part of the spectrum. } 
\label{fig:comb_spec}
\end{figure*} 

\begin{table}[t]
    \centering
    \caption{Photometry for AzTEC-3 extracted from {\it HST}  images using SExtractor.}
    \label{tab:az3_phot}
    \begin{tabular}{l l} \hline \hline
        Filter & Magnitude \\
         & [$\rm mag_{AB}$] \\ \hline
        F606W & $28.89 \pm 0.30$ \\
        F814W & $27.17 \pm 0.10$ \\
        F105W & $24.92 \pm 0.03$ \\
        F125W & $24.68 \pm 0.02$ \\
        F160W & $24.52 \pm 0.02$ \\
        \hline
    \end{tabular}
\end{table}

\section{Discussion} \label{sec:discussion}

\subsection{Magnetic fields} \label{subsec:Mag}

The mechanisms for the predicted polarization of the \cii fine
structure line has been named  ground state alignment (GSA)
\citep{Yan2006, Yan2012, Zhang2015}. GSA occurs due to the interaction of an
anisotropic radiation field with atoms or ions with fine- or
hyperfine-structures. A magnetic field induces precession, which causes
the atom or ion to align with the magnetic field. As a result, the
emission or absorption of the spectral lines becomes polarized and the
polarization direction reflects the direction of the magnetic
field. The atoms align predominantly at their ground state
level. Due to the low emission and absorption rates involved in
these transitions, and hence their long life spans, already weak fields
($B>10^{-15}$~G) can cause alignment.

Predictions for the level of \cii polarization based on the GSA
effect have been made for \cii emission near galactic star-forming
regions \citep{Zhang2018}. Near the strong anisotropic radiation field produced
from the regions, \cii polarization up to $\sim30\%$ could be
expected. Since the radiation field in starburst galaxies and AGN
hosts are similarly energetic, comparable levels of polarization could have
been expected in early star-forming galaxies and AGNs \citep{Zhang2018}. Our observations do not reveal any linear polarization of \cii with $3\sigma$ limits to the polarization percentages at the peak of emission <5\% and for the integrated emission <1\%. 

There are two possible explanations for
such low or non-existent levels of polarization. Firstly, the GSA
prediction might not be relevant in conditions where the \cii
emission originates in the observed SMG and quasar, because the
anisotropic radiation field is not sufficient to cause significant
polarization. Alternatively, if the magnetic field is sufficiently
irregular or has a large turbulent component within our resolution
elements, beam depolarization will reduce the polarization
fraction. Our angular resolution corresponds to $\sim3.1$\,kpc and
$\sim4.7$\,kpc for BRI0952 and AzTEC-3, respectively. These scales are
not significantly larger than those where structured fields are
observed in nearby galaxies \citep[e.g.,][]{Beck2015, Lopez2021b}, but if the magnetic
field follows spiral arms or warped disks, like in the case of
Centaurus A \citep{Lopez2021a}, depolarization could still be large. Without a
better estimate of the expected linear polarization fraction in SMGs
and around quasars, we cannot use our limits to provide meaningful
constraints on the level of depolarization, and thus the level of
structure in the magnetic field. However, unless the GSA mechanism is much
less effective than expected, it appears unlikely that a strong
ordered magnetic field exists at the kiloparsec scale in the two sources in this paper.

\subsection{Infrared luminosity} \label{subsec:LIR}

We obtain an $L_{\rm IR}$ estimate for AzTEC-3 by integrating our SED from Bagpipes from 8-1000 $\mu$m of $L_{\rm IR} = (7.3 \pm 0.2) \times 10^{13}\,\rm L_{\odot}$. Although the central source is blended with surrounding companions, we anticipate that the uncertainty caused by the companion's IR luminosity will be small and we do not add additional errors to the $L_{\rm{IR}}$ of AzTEC-3. This value is consistent with that reported by \citet{Capak11}, as they provide a wide range of possible $L_{\rm IR}$ values ranging from $(2.2-11) \times 10^{13}\,\rm L_{\odot}$ for the 8-1000\,$\mu$m wavelength range. \citet{Riechers14} reported a $L_{\rm FIR}$ that is a factor of $\sim6.5$ times lower than our reported $L_{\rm IR}$, which is outside the range of error typically ascribed to differences between $L_{\rm IR}$ and $L_{\rm FIR}$ \footnote{In this analysis, we assume $L_{\rm FIR} \sim 0.75 \times L_{\rm IR,}$ following \citet{Decarli17}.} \citep{Carilli13}. We suggest that discrepancies between these two reported values, apart from the utilized wavelength range, derive from the number of data points used for each fit. \citet{Riechers14} used a higher number of data points than in our SED fitting; however, both analyses lack IR coverage, making it difficult to obtain high levels of accuracy, and thus $L_{\rm FIR}$ or $L_{\rm IR}$, from the SED fit. 

To determine a lower limit on the IR luminosity of the companions of AzTEC-3, we used a modified blackbody approximation (e.g., \citet{Knudsen03}, Equation 2), assuming a temperature of 45\,K and $\beta = 1.7$ \citep[typical values for high-redshift sources;][]{Beelen06, Dunne11, Carniani19}. We utilized this approach rather than an SED fit as the SED of the companions is poorly sampled, especially in the far-IR, and the accuracy of sub-mm photometry is affected by the blending of the continuum emission from the central SMG. Calculated lower limits are provided in Table \ref{tab:derived_props}.

A similar fit for the IR luminosity was performed for BRI0952, integrating the SED between 8 and 1000\,$\mu$m. We assume the same approximation as for AzTEC-3: the effect of the companion sources on the SED and $L_{\rm{IR}}$ will be negligible due to their faintness in comparison to the quasar. Specifically, we find that the IR luminosity due to star formation is $L_{\rm{IR_{SF}}} = (7.66 \pm 2.49) \times 10^{12} \,\rm L_{\odot}$ and the IR luminosity due to AGN is $L_{\rm{IR_{AGN}}} = (2.23 \pm 0.72) \times 10^{14} \,\rm L_{\odot}$. We note that the error on the IR luminosity comes from the uncertainty on the lensing factor. We find that $L_{\rm{IR_{SF}}}$ is a factor of $\sim 5$ higher than that reported by \citet{Gallerani12}. We suggest that this is for two main reasons: (i) the methodology used for fitting - \citet{Gallerani12} scaled a template SED to the 870$\mu m$ continuum flux, while our model fits based on a number of photometric values, increasing the accuracy of the fit - and (ii) the wavelength range used in fitting \citep[the range 42-122\,$\mu m$ is used in][]{Gallerani12}. 

\begin{table*}[t]
    \centering
    \caption{Derived properties for sources. $L_{\rm IR}$ refers to the integrated IR luminosity from 8-1000 $\mu$m for BRI0952 (using $L_{\rm{IR_{SF}}}$) and AzTEC-3 and to a modified blackbody approximation for the companion sources. $\rm SFR_{IR}$ is the SFR derived from the $L_{\rm IR}$, $\rm SFR_{[CII]}$ refers to the SFR derived from the relation given in \citet{DeLooze14} for starburst galaxies. 
    $\dot{M}_{\rm out}$ gives the mass outflow rate, and $M_{\rm dyn}$ is the dynamical mass. Upper limits are provided for inferred properties for Gal-S, Comp-N, and Comp-SW as we do not detect these sources in continuum emission. The luminosities reported for BRI0952 are corrected for magnification. }
    \begin{tabular}{l c c c c c c c} \hline \hline 
        Name & $L_{\rm [CII]}$ & $L_{\rm IR}$ & $\rm SFR_{IR}$ & $\rm SFR_{\rm [CII]}$ & $L_{\rm [CII]}/L_{\rm IR}$ & $\dot{M}_{\rm out}$ & $M_{\rm dyn,[CII]}$ \\
        & [$10^{9}$\,L$_\odot$] & [$10^{12}$\,L$_\odot$] & [M$_{\odot}$\,yr$^{-1}$] & [M$_{\odot}$\,yr$^{-1}$] & $10^{-3}$ & [M$_{\odot}$\,yr$^{-1}$] & [$10^{10}$\,M$_{\odot}$] \\ \hline
        
        BRI0952 & $2.46 \pm 0.23$ & $7.66 \pm 2.49$ & 770 & 247 & 0.37 & $98 \pm 19$ & 0.90 \\
        
        Comp-N & $0.39 \pm 0.04$ & <0.2 & <21 & 34 & <2.0 & - & 0.5 \\
        
        Comp-SW & $0.19 \pm 0.03$ & <0.2 & <21 & 16 & <1.0 & - & 0.4 \\ \hline
        
        AzTEC-3 & $9.3 \pm 0.80$ & \azir & 7340 & 809 & $0.12 \pm 0.01$ & $238 \pm 30$ & 6.5\\
        
        LBG-3 & $0.63 \pm 0.05$ & $0.40 \pm 0.5$ & $40 \pm 46$ & 55 & $1.6 \pm 0.14$ & - & 0.95 \\
        
        Gal-S & $0.43 \pm 0.04$ & $<0.1$ & $<12.0$ & 37 & $<3$ & - & 0.83 \\ 
        
        Gal-SW & $0.12 \pm 0.01$ & $0.15 \pm 0.2$ & $15 \pm 23$ & 11 & $0.85 \pm 0.13$ & - & 1.6 \\ 
        
        \hline
    \end{tabular}
    \label{tab:derived_props}
\end{table*}

\subsection{Star-formation rate} \label{subsec:sfr}

We calculated the SFR of BRI0952 and AzTEC-3, along with their surrounding sources, using different methods. By assuming a Chabrier IMF we can infer the SFR of each source via the following relation: SFR $\sim 10^{-10} L_{\rm{IR,}}$ where $L_{\rm IR}$ is given in $L_{\odot}$ \citep{Carilli13}. This yields an SFR of 770 M$_{\odot}$\,yr$^{-1}$ for the quasar BRI0952 using $L_{\rm{IR_{SF}}}$ and $\sim 7340$ M$_{\odot}$\,yr$^{-1}$ for AzTEC-3 (significantly higher than that reported by \citet{Riechers14} of 1100 M$_{\odot}$\,yr$^{-1}$). 

These extremely high values are similar to those reported for other peculiar sources in the high-redshift universe. \citet{Daddi09} reported an SFR of $> 1000 \rm M_{\odot}/yr$ in the SMGs GN20 and GN20.2 and notes that there is no evidence of AGN activity. Similarly, SFRs of $\geq 1000$ M$_{\odot}$\,yr$^{-1}$ have been reported in HDF850.1 \citep{Walter12}, AzTEC-1 \citep{Yun15, Sharda19}, HFLS3 \citep{Robson14, Cooray14}, and BRI1202 \citep{Carilli13_2}. 

Following the same procedure we infer the SFRs of the companions of AzTEC-3 from the modified blackbody $L_{\rm{IR}}$ fit, reported in Table \ref{tab:derived_props}. The companions exhibit significantly lower SFRs than AzTEC-3, although we assume that these values provide a lower limit for the companions. We find comparable SFRs to those reported for the quasar companions in \citet{Neeleman19}. 

We derive SFRs through an additional method using the \cii-SFR relation for starburst galaxies from \citet{DeLooze14} (provided in Section \ref{subsec:BRI_cii}), which are given in Table \ref{tab:derived_props}. This results in a similar distribution to the above method, demonstrating the significant discrepancy between the SFR of the companions and that of the central sources in both fields. 

We caution that using \cii as a means of inferring SFR may be inaccurate as the \cii emission is likely tracing other processes in addition to star formation within the galaxies. The \cii emission could be tracing more extreme processes or neutral gas \citep{Pavesi18} within the galaxy. If this is the case in AzTEC-3 or BRI0952, this could also contaminate the measurement of the SFR from the \cii emission. We further note the degeneracy inherently present in determining a SFR from a galaxy's \cii luminosity due to the \cii versus FIR deficit, which is explored below. 

\subsection{[CII] Deficit} \label{subsec:CIIDeficit}

\begin{figure}[h]
    \centering
    \includegraphics[width = 1.0\linewidth]{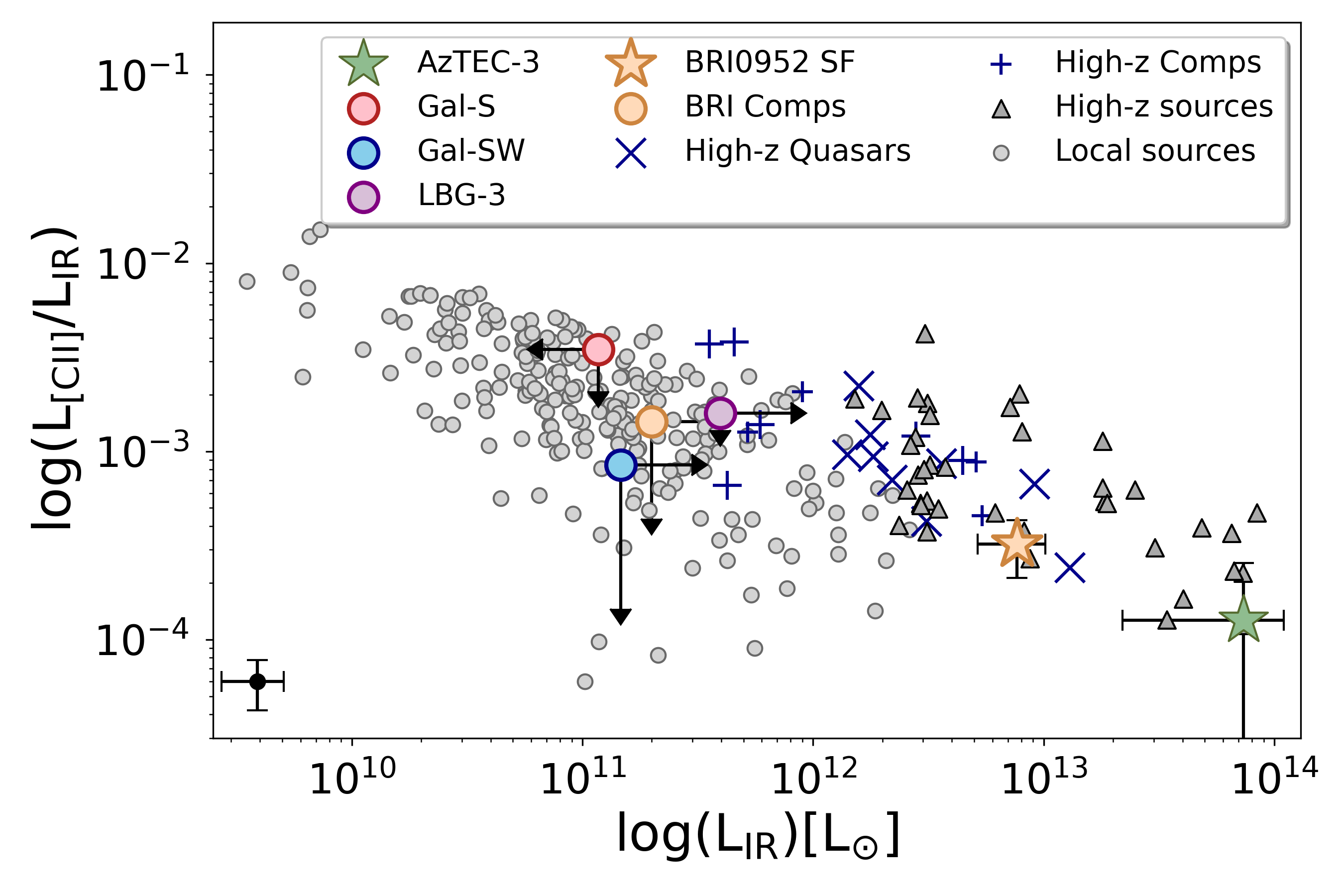}
    \caption{$L_{\rm [CII]}$/$L_{\rm IR}$ ratio as a function of $L_{\rm IR}$ for local and high-redshift sources. Local ($z<1$) sources are taken from \citet{DiazSantos13}. High-redshift sources are from \citet{DeLooze14} and \citet{Gullberg15}. Literature results for high-redshift quasars and companion sources are taken from \citet{Decarli17} and \citet{Neeleman19} (for these we take an average of the reported $L_{\rm IR}$). We note that we only consider the $L_{\rm [CII]}$/$L_{\rm IR}$ as an upper limit for all companion sources. The error bar we used for BRI0952 comes from the uncertainty on the lensing factor on $L_{\rm IR}$. The error bar on the $L_{\rm IR}$ and hence $L_{\rm [CII]}$/$L_{\rm IR}$ ratio of AzTEC-3 is taken to be the range of $L_{\rm IR}$ values provided by \citet{Capak11}. For data points taken from other papers, we assumed $L_{\rm FIR} \sim 0.75 \times L_{\rm IR}$ \citep[following][] {Decarli17}, but added an additional indicative error bar to the bottom left of the plot indicating the conservative estimate $\sim30\%$ due to this assumption \citep{Carilli13}.} 
    \label{fig:CII_FIR}
\end{figure}

The \cii line is well known to exhibit a deficit with increasing IR luminosity \citep[e.g.,][]{DiazSantos13, Gullberg15, DiazSantos17, Gullberg18, Lagache18}. This has been heavily investigated at high redshift \citep[e.g.,][]{Stacey10, Wang13, Gullberg15, Decarli17, Gullberg18, Lagache18, Neeleman19}, with many proposed explanations including the physical scale of star formation, \cii saturation, optical depth effects, increased dust grain charge in PDRs and the ISM, and AGN activity \citep{Casey14}. Some studies at high redshift have suggested that the lowest ratios occur preferentially in AGN host galaxies \citep{Stacey10}. This may not be exclusively limited to \cii emission, as other fine structure lines such as [O\,{\sc i}], [O\,{\sc iii}], [N\,{\sc ii}], and [N\,{\sc iii}] have been found to exhibit this deficit as well \citep[e.g.,][]{GarciaCarpio11, Decarli12, Farrah13}. 
We plot the $L_{\rm [CII]}/L_{\rm{IR}}$ ratios as a function of the IR luminosity for the sources in our sample, as well as other low- and high-redshift galaxies in Figure \ref{fig:CII_FIR}. The ratios we find for BRI0952 and AzTEC-3 are similar to those found in HFSL3 \citep{Riechers13} and the ratios reported for two of the four quasars studied in \citet{Decarli17}. 

For the companion galaxies, we also investigated the $L_{\rm [CII]}/L_{\rm{IR}}$ ratio, though we note that the errors are very large due to the additional uncertainty caused by the deblending of emission from the central bright source.  As the IR luminosity is seen as a lower limit (see Section ~\ref{subsec:LIR}), the ratio can be treated as an upper limit.  The resulting values are consistent with those of local star-forming galaxies \citep{DiazSantos13} and are higher than those reported for high-redshift companion sources detected in \cii \citep[e.g.,][]{Carilli13_2, Decarli17, Neeleman19}. 

A possible explanation for the ratios observed in AzTEC-3 and BRI0952 is that of \cii saturation \citep[e.g.,][]{Munoz16, Gullberg18}. These sources are both hosts to extreme star formation and, in the case of BRI0952, AGN activity. If the temperatures in the majority of the environments in which \cii is produced exceed the ground state temperature (92\,K), we can expect this line to saturate and other fine structure lines to become the primary coolants of the ISM. If this is the case for these two sources, it could explain the deficits observed in both. Observations of other fine structure lines in these sources could provide further clues to the origin of the deficit. 
\subsection{Outflows and turbulence} \label{subsec:outflows}

As shown in Sections \ref{subsec:Az3_cii} and \ref{subsec:BRI_cii}, the \cii emission line profiles of the bright target sources, namely BRI0952 and AzTEC-3, are better fit when including an additional broad Gaussian function. The presence of broad, higher velocity wings in the line profiles are often interpreted as an indication of high-velocity outflows. We note that this need not be a unique interpretation; however, as this is a common analysis that has been done in many previous works \citep[e.g.,][]{Feruglio10,Aalto12,Maiolino12,Cicone14,Cicone15, Gallerani18,Bischetti19,Stanley19,Ginolfi20}. We pursue this below. 

We used the luminosity of the broad component of the \cii line in both AzTEC-3 and BRI0952 to infer a mass outflow rate using the equation from \citet{HaileyDunsheath10} to calculate the mass of the outflow. Similarly to the method employed by \citet{Maiolino12}, \citet{Cicone15}, and \citet{Stanley19}, we assumed $\mathrm{X_{C^+} = 1.4 \times 10^{-4}}$, $T = 200$\,K, and $n \gg n_{\rm crit}$ - typical values for PDRs. We estimated the velocity of the outflow by assuming a constant outflow rate of $v_{\rm outf} = 0.5 \times \mathrm{FWHM_{broad}}$ over a region with a radius of $\mathrm{R_{out}}$, allowing us to calculate the outflow rate using $\dot{M}_{\rm out} = M_{\rm out} \times v_{\rm out}/R_{\rm out}$. 

We determined the outflow radius as the major axis of the extent of the \cii emission for AzTEC-3 ($\sim 7.5$\,kpc) and BRI0952 ($\sim 2.3$\,kpc); for BRI0952, this measurement was taken using the major axis through both the image plane top and bottom images combined, and corrected for lensing magnification \footnote{The lensing magnification factor we utilized to correct the extent was $\mu = 3$.}. We investigated alternative methods for determining the mass outflow rate below. We find outflow rates of $\dot{M}_{\rm out} = 238 \pm 30 \mathrm{M_{\odot}/yr}$ for AzTEC-3 and $\dot{M}_{\rm out} = 98 \pm 19 \mathrm{M_{\odot}/yr}$ for BRI0952. Although these values are lower than those found for other individual detections of quasar-driven outflows \citep[assumed to be the case for BRI0952; e.g.,][]{Feruglio10, Sturm11, CanoDiaz12, Maiolino12, Cicone14, Feruglio15}, they are in good agreement with outflow rates estimates through stacking analyses \citep[e.g.,][]{Gallerani18,Stanley19,Ginolfi20}. The latter samples are comprised of more `normal' high-redshift sources, so the validity of the comparison is tenuous and the methodology used is discussed at the end of this section. 

As noted above, there are different methods for disentangling a potential contribution of a high-velocity outflow to the line profile, and we note that the velocity profile of the outflow need not be described by a single Gaussian. As an alternative, we determined the strength of the outflow by utilizing only the flux not accounted for in a single Gaussian fit to the \cii emission; effectively subtracting the single component from the broad one. This yields a significantly lower mass outflow rate, which we treated as a lower limit for BRI0952 and AzTEC-3. Using this method, we find mass outflow rates of $28 \pm 8 \rm M_{\odot}/yr$ for AzTEC-3 and $12 \pm 4 \rm M_{\odot}/yr$ for BRI0952. We attribute the discrepancy between the two methods, at least in part, to be the effect of the companion sources on the \cii line emission profile. If the emission from the central source is blended with that of the companions, it is possible that the double-Gaussian fit (and thus the broad complex emission line profiles) are simply an artifact caused by this blending.

We caution that the differential lensing of the quasar may be contributing to the high-velocity wings. To this end, for an additional investigation into the mass outflow rate of BRI0952 we used the spectra extracted from Img-N (as mentioned in Section \ref{subsec:BRI_cii} and shown in Figure \ref{fig:BRI_spec}). We use a radius of $0.35''$ ($\sim2.38$kpc), corrected using the same lensing factor as described above), $\rm FWHM_{broad}$ = $543$\,km\,s$^{-1}$, and $L_{[\rm C\,{\sc II}]_{broad}} = 0.22 \times 10^{9}$\,L$_{\odot}$, resulting in an outflow of $\dot{M}_{\rm out} = 74 \pm 19\,\mathrm{M_{\odot}/yr}$. The exact impact of the gravitational lensing, and in particular differential lensing, is challenging to estimate, and further modeling based on higher resolution and higher sensitivity data across multiple wavelengths would be needed. We note that systematic errors on the mass outflow rate estimates, along with the systematic errors on other quantities used for comparison (e.g., SFR), are likely to dominate over the effect of the differential lensing. 
An additional important consideration is the origin of this broad component, which we interpret as an outflow above; for example, we ask ourselves whether the broad component is tracing high-velocity gas outflowing from AzTEC-3 or an interaction between the SMG and the companion Gal-S. This is further discussed in section \ref{subsec:environment}. 

\begin{figure}[h]
    \centering
    \includegraphics[width = 1.0\linewidth]{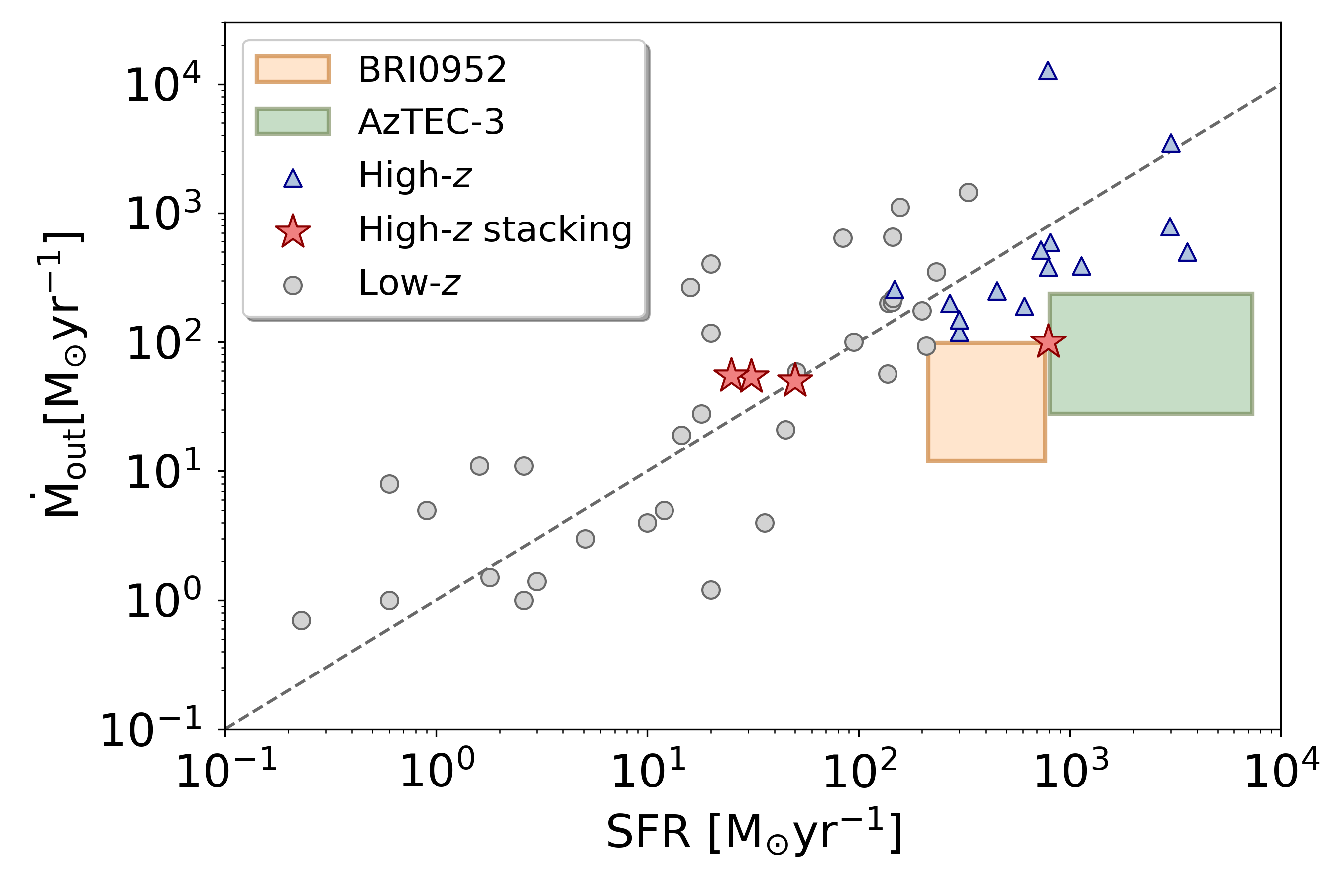}
    \caption{Mass outflow rate versus SFR for our objects and low- and high-redshift galaxies, including stacking approaches at high redshift. The green and orange rectangles represent the range of possible values for AzTEC-3 and BRI0952, respectively. 
    The low-$z$ comparison sample is taken from \citet{Fluetsch19}, high-redshift direct observations are taken from \citet{Maiolino12, George14, Feruglio17, Brusa18, HerreraCamus19, Jones19, Spilker20_b, Butler21}, and high-redshift stacking averages are taken from \citet{Gallerani18, Bischetti19, Ginolfi20}. For \citet{Ginolfi20}, we plot the high-SFR sample and the median-SFR sample in their stacking methodology as separate points. We utilized an average of the mass outflow rates if a range is provided for an object.}
    \label{fig:sfr_mout}
\end{figure}

We show the mass outflow rate estimates as a function of SFR in Figure \ref{fig:sfr_mout}, together with similar estimates for low- and high-redshift sources. With star-formation-rate estimates that are $\sim \rm or > 1000$\,M$_\odot$\,yr$^{-1}$ for both BRI0952 and AzTEC-3, both sources are seen in a similar region to other results for high-redshift galaxies. For both sources, the mass outflow rate estimates are on the lower side of the average; however, we note that the large uncertainties in the $\dot{M}_{\rm out}$ and SFR do not allow for additional interpretation.         

Outflows are generally studied through different probes, including both emission from high-velocity outflows as well as absorption line studies. In terms of high-velocity outflows, detections have been published using different lines, including CO, [C\,{\sc ii}], and [O\,{\sc iii}]$\lambda\,5007$ at low and high redshift \citep[e.g.,][]{Cicone14, Cicone15, Carniani16, Brusa18}. So far, only a few robust detections from single $z>4$ quasars using \cii have been published \citep{Maiolino12,Carilli13_2}. Studies using stacking analyses for larger samples of $z\sim 6$ quasars provide conflicting results \citep[e.g.,][]{Decarli18,Stanley19,Bischetti19}, ranging from non-detection to claimed detections. In terms of star-forming galaxies, stacking of \cii for $z=4-6$ galaxies in the ALPINE survey revealed a high-velocity outflow component, finding mass outflow rates that are consistent  with our results \citep[e.g.,][]{Ginolfi20}. The use of broad \cii emission as a means of detecting outflows was called into question by \citet{Spilker20} following their non-detection of broad wings associated with a sample of dusty star-forming galaxies with clear OH outflow absorption features. It is thus possible that the broad wings we detect are dominated by emission from the companion sources around AzTEC-3 and BRI0952, where the latter could also be affected by differential lensing magnification. In addition, recent studies have reexamined previous results, indicating the need for a broad component in the fitting of \cii spectra and found these components superfluous \citep[e.g.,][]{Meyer22}; therefore, we caution that the true nature of the broad component cannot be confirmed as an outflow in BRI0952 and AzTEC-3 without additional outflow tracers detected in these galaxies.
\subsection{Environment} \label{subsec:environment}

The impact of the environment in which massive galaxies evolve at high-redshift remains an open question. These two systems present extreme situations in which to study the effects of faint companion sources in the early universe.  

The AzTEC-3 system, with the presence of multiple close companions, detected either in \cii or in optical or continuum observations, provide an exceptional laboratory to study the effect of close companions near intense starbursts in the early Universe. There are three companion galaxies within a projected distance of $\sim18$\,kpc from the central SMG, and an additional system of possibly merging galaxies located at a projected distance of $\sim 95$\,kpc from the central source \citep{Riechers14}, though the latter is not covered by our observations. We detect a bridge-like structure between the companion galaxy Gal-S and AzTEC-3, suggesting the possible occurrence of a gas-exchange between the two galaxies extending over $\sim 12$\,kpc. This "gas bridge" between Gal-S and AzTEC-3 is very similar to that observed by \citet{DiazSantos18} between the Hot DOG W2246 and a companion, although only about half as large in spatial extent. Further investigation of this requires higher resolution ALMA data and an improved method for isolating and subtracting the emission from the central source. As mentioned in Section \ref{subsec:aztec3}, we detect a velocity gradient between LBG-3 and AzTEC-3, indicating an additional possible interaction. The detections of Lyman-$\alpha$ between these two galaxies in \citet{Guaita22} also indicate an interacting system. 
 
In the field of BRI0952, Comp-N and Comp-SW are detected in \cii with no further sources detected out to a projected distance of 62\,kpc (corresponding to the radius of the primary beam). Also, no additional companions are seen at other wavelengths, most notably the archival {\it HST} data. Due to the non-detection of the companions in current {\it HST} and {\it Herschel} imaging, an analysis of comparable level to that of AzTEC-3 is currently not feasible. We note the possibility that one or both of Comp-N and Comp-SW could be in the process of merging with BRI0952. If this is the case, it may suggest that we are observing the quasar in a post-starburst state in which recent galaxy interactions and ongoing mergers have triggered extreme star formation and AGN activity. In order to investigate this possibility, higher-resolution data of multiple emission lines combined with a robust source plane reconstruction would be needed; this is beyond the scope of this paper. For the BRI0952 companion sources, we see no clear signs of gas-bridge-like structures (as were seen for the AzTEC-3 companion galaxies). We also note that an alternative interpretation of either of the companions Comp-N and Comp-SW could be that they represent an extended substructure in the gas distribution. If that is the case, it would likely indicate the presence of merger activity, as the gravitational forces from minor or major merger interactions could cause a more complex gas distribution \citep[e.g.,][]{Konig14,Harada18, Konig18, Young21}.

These two systems seem to be examples of long-sought-after, theorized, typical, over-dense environments of massive sources with numerous faint companions in the high-redshift Universe. The companion sources of both systems contribute less than 10\% to the total \cii emission, and even less to the total IR luminosity between 8-1000\,$\mu m$. Other systems observed in recent years have also been found to have companion sources; however, most of these companions have luminosities comparable to that of the central SMG or quasar host galaxy  \citep[e.g.,][]{Clements09,Carilli13_2, Robson14,Fogasy17,trakhtenbrot17,wardlow18,DiazSantos18,Neeleman19,Fogasy21,Bischetti21}. The detection of faint companion sources in these two fields, together with the results of W2246$-$0526 and BRI1202$-$0725 \citep{Carilli13_2,DiazSantos18} and other such systems, are increasing the sample enabling investigation of the role of less massive companion sources on massive galaxy evolution.

Theoretical predictions from semi-analytical model simulations suggest that $22\%$ of quasars should have at least one companion galaxy with stellar masses $>10^8$\,$\mathrm{M_{\odot}}$ \citep{Fogasy17}. Additionally, studies show that minor mergers, especially in the high-redshift Universe, are common. \citet{Kaviraj15} utilized the Horizon-AGN hydrodynamical cosmological simulation to show that by $z \sim 1$ all massive galaxies ($>10^{10}\,\rm M_{\odot}$) have undergone a major or minor merger, and that minor mergers (those with a mass ratio > 4:1) are around $2.5\times$ more frequent than major mergers between $1 < z < 4$. Their work also suggests that major mergers are not the dominant source of star-formation enhancement at high redshift \citep[see Figure 5 in ][]{Kaviraj15}. This is indicative of the need for minor mergers as fuel providers for high-redshift galaxies and is especially important for extreme SMGs hosting maximum starbursts such as AzTEC-3. The lack of current detections of smaller companion sources is likely due, in part, to the long integration time required to observe them. 

In order to categorize these systems as possible minor mergers (be it progenitors or ongoing processes), we calculated the mass of the central source using a virial mass estimator following the procedure used by \citet{Riechers14}. We find the dynamical mass inferred from the \cii emission to be $\sim 6.5\times10^{10}$\,M$_{\odot}$ for AzTEC-3 and $\sim 0.90 \times10^{10}$\,M$_{\odot}$ for BRI0952 (lensing corrected). We used the same method to compute the masses of the companions, reported in Table \ref{tab:derived_props}. In the AzTEC-3 system, we find that the companions have dynamical masses of $\leq 4 \times$ that of AzTEC-3. This would classify these companions as minor mergers should they interact with the central source; we already observe signs of this for AzTEC-3 and its companions in the form of the gas bridge between these objects. These companion sources are significantly less massive than those reported by \citet{Neeleman19}. For the companions of BRI0952 we find the companions have masses $\sim1.5 -2 \times$ smaller than that of the quasar. These values for the companions are closer to the values found for the quasar companions in \citet{Neeleman19}. We note that the companion masses are likely overestimated as we do not correct for lensing along the major axis of the companions used for the virial mass estimates due to uncertainties in our lensing model, and therefore the above estimate can be seen as an upper limit.

The effect of the companions on massive sources remains to be seen. If these faint companions are dynamically interacting in some manner, such as providing gas to the central sources, this could supply the needed materials for the extreme star formation occurring in these systems. Additionally, as suggested by \citet{McGreer14}, the possibility of mergers occurring on a relatively fast timescale as a short transitional phase could drastically limit our ability to obtain high number density observations of similar systems. We further suggest that the increased resolution now possible with ALMA will allow for increased detections of SMG and quasar systems with numerous and faint surrounding objects.

\section{Conclusions} \label{sec:conclusions}
In this paper, we present observational results of \cii emission from BRI0952 and AzTEC-3, along with respective companion sources. Our results lend credibility to the paradigm of major and minor mergers in the early Universe as progenitors for the massive galaxies we see in studies of the local Universe. We summarize our conclusions below. 

   \begin{enumerate}
      \item We detect \cii emission in the lensed quasar BRI0952 at $z \sim 4.433$ and the SMG AzTEC-3 at $z \sim 5.3$. We report serendipitous detections of \cii emission from two previously unreported companion sources around BRI0952 and new detections of \cii emission from three companions surrounding AzTEC-3. These companions are each located within 3$\,''$ (18\,kpc) of the central source. 
      
      \item We present a full-polarization analysis of the \cii emission lines for both main targets. No polarization was detected, and upper limits are provided. The results suggest that strong ordered magnetic fields are unlikely to exists at the kiloparsec scale in the two studied sources, unless ground state alignment is a less effective mechanism than expected. 
      
      \item We constructed a new lensing model for BRI0952 using {\sc Visilens} \citep{Spilker16}, yielding a lensing magnification factor of $\mu \sim 4$ for the quasar and insignificant lensing magnification of the two companion sources. Our model suggests that differential lensing is occurring across the surface of BRI0952 in both \cii and continuum emission. This difference is likely insubstantial (or within errors) for our purposes, but it is important to keep it in mind when considering the physical properties of the lensed images. 
      
      \item The inferred SFR from the IR luminosity of both the central SMG AzTEC-3 and the quasar BRI0952 suggest that both sources harbor starbursts of $\sim \rm or > 1000$ solar masses per year. 
      
      \item The central SMG AzTEC-3 and a companion galaxy (Gal-S) in the field show evidence of an interlinking gas bridge. Although we do not find a strong velocity gradient across the central source, we suggest that this bridge may be indicative of an ongoing gas-exchange process or merger.  
      
      \item The \cii line profiles for both central sources exhibit complex broad features indicating the possible presence of outflows. The mass outflow rates of both BRI0952 and AzTEC-3 are similar to results for high-redshift galaxies; any discrepancies we find are likely symptomatic of large uncertainties on both the mass outflow rate and the SFR. 
      
      \item The outflow features, combined with the observed gas-bridge structure between AzTEC-3 and its southern companion (and possibly others) and velocity-gradients between BRI0952 and companions, suggest both are interacting systems. The extent of this interaction is unknown, but if both systems are either entering or exiting a merger phase, this could explain the extreme star formation events occurring in both.
   \end{enumerate}

Growing evidence in recent years suggests that overdense regions leading to major and minor mergers are the progenitors of the massive galaxies we see in the local Universe. The improved resolution possible with ALMA will allow for the increased detection of companion galaxies in high-redshift environments, allowing us to explore the credibility of mergers as means of creating the extreme SFRs observed in these objects. 

\begin{acknowledgements}
We thank the anonymous referee for their helpful comments and insights. Kiana Kade acknowledges support from the Nordic ALMA Regional Centre (ARC) node based at Onsala Space Observatory. The Nordic ARC node is funded through Swedish Research Council grant No 2017-00648. Kirsten Knudsen acknowledges support from the Swedish Research Council and the Knut and Alice Wallenberg Foundation. BG acknowledges support from the Carlsberg Foundation Research Grant CF20-0644 `Physical pRoperties of the InterStellar Medium in Luminous Infrared Galaxies at High redshifT: PRISMLIGHT'. Sabine König gratefully acknowledges funding from the European Research Council (ERC) under the European Union’s Horizon 2020 research and innovation programme (grant agreement No 789410).
This paper makes use of the following ALMA data: 2018.1.01536.S. ALMA is a partnership of ESO (representing its member states), NSF (USA), and NINS (Japan), together with NRC (Canada), MOST and ASIAA (Taiwan), and KASI (Republic of Korea), in cooperation with the Republic of Chile. 
This paper makes use of the following {\it HST} projects: 13641 (PI Capak), 9822 (PI COSMOS24-21), and 13384 (PI Riechers), 8268 (PI Impey). Based on observations made with the NASA/ESA Hubble Space Telescope, and obtained from the Hubble Legacy Archive, which is a collaboration between the Space Telescope Science Institute (STScI/NASA), the Space Telescope European Coordinating Facility (ST-ECF/ESA) and the Canadian Astronomy Data Centre (CADC/NRC/CSA)
This research made use of APLpy, an open-source plotting package for Python (Robitaille \& Bressert 2012).
\end{acknowledgements}

\bibliographystyle{aa}
\bibliography{aanda}

\begin{appendix}

\section{Continuum-subtracted spectra}

\begin{figure*}
    \centering
    \includegraphics[width = 0.8\linewidth]{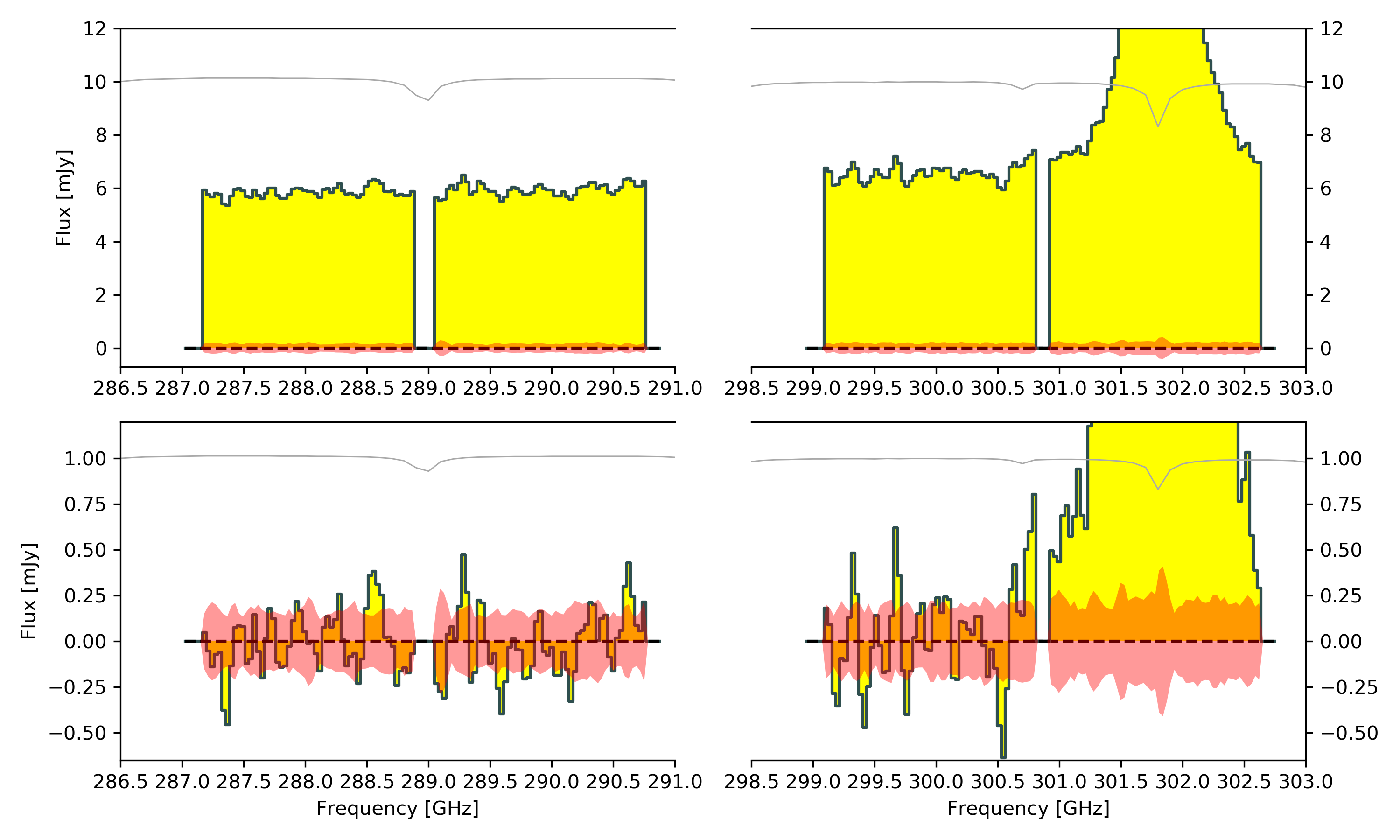}
    \caption{Pre-continuum-subtracted (top) and post-continuum-subtracted (bottom) spectra of AzTEC-3 extracted from the same region as the spectra (black circle shown in panel 1 of Figure \ref{fig:Az3_mom0} using a fit on the order of 1 and encompassing frequencies up to 300GHz. The red region represents the $1\sigma$ rms of the image and the gray line represents the atmospheric transmission at 1.0 mm PVW. The possible additional \cii wing feature mentioned in Section \ref{sec:obs} can be seen in the 300.5-301\,GHz frequency range. }
    \label{fig:Az3_contsub}
\end{figure*}

\begin{figure*}
    \centering
    \includegraphics[width = 0.8\linewidth]{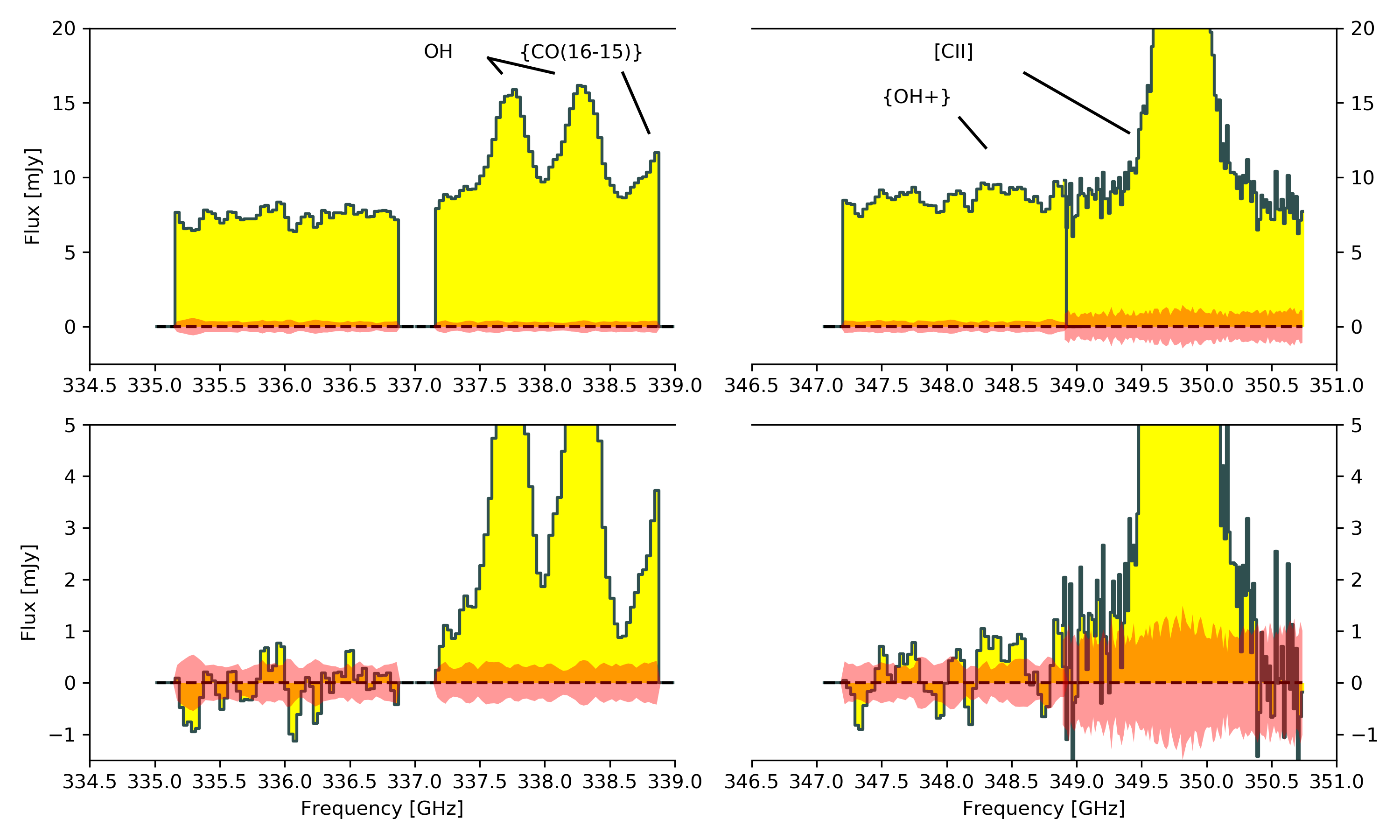}
    \caption{Pre-continuum-subtracted (top) and post-continuum-subtracted (bottom) spectra of BRI0952 extracted from the same region as the spectra (black circle shown in panel 1 of Figure \ref{fig:BRI_mom0} using a fit on the order of 1 and encompassing frequency ranges from $\sim335.3-336.9$, $347.1-347.7$, $348.97-348.98,$ and $350.45-350.67$\,GHz. The red region represents the $1\sigma$ rms of the image. We do not show the atmospheric transmission as there are no relevant features in the frequency range. We note that the spectral window of the \cii line has a higher spectral resolution (see Section \ref{sec:obs}), resulting in a higher noise level per channel compared to other spectral windows.}
    \label{fig:BRI_contsub}
\end{figure*}

Presented in this appendix are the figures showing the total spectra compared with the continuum-subtracted spectra for each source (see Figures~\ref{fig:Az3_contsub} and \ref{fig:BRI_contsub}). 

\section{Lensing model for BRI0952}

\begin{figure*}
    \centering
    \includegraphics[width = 0.8\linewidth]{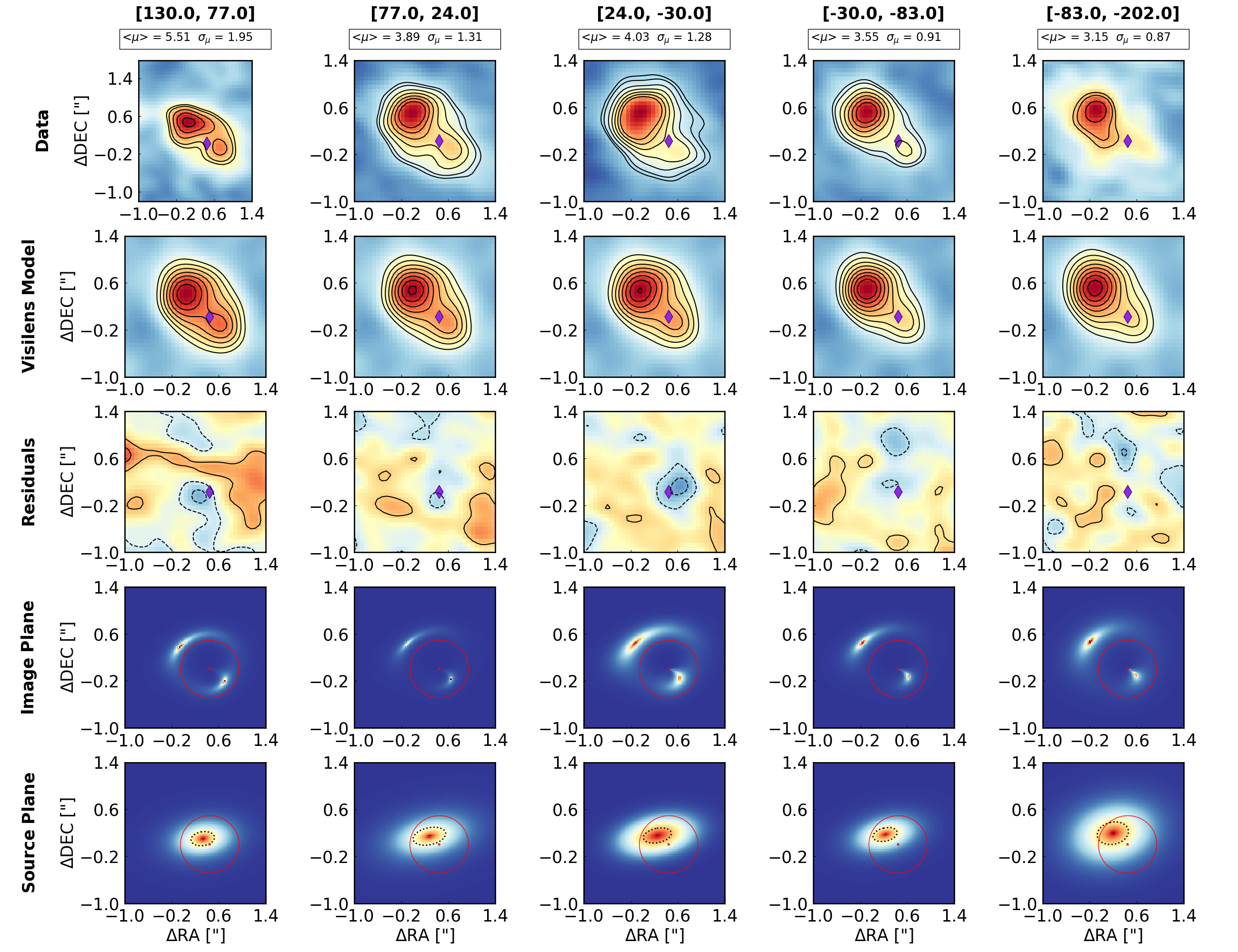}
    \caption{Image and lens models for BRI0952 for the velocity range shown at the top of each row. The diamond in the first three rows shows the best-fit position of the lens. Panels from top to bottom: Row 1 shows the ALMA band-7 dirty image with contours at 4, 6, 8, 10, 12, 14, and 16 $\sigma$ levels and the magnification value found for that bin; row 2 shows the image plane model from {\sc Visilens} with the same contour levels as row 1; row 3 shows the residuals after the subtraction of the {\sc Visilens} model, and contours are at -3, -2, -1, 1, and 2 $\sigma$ levels; row 4 shows the image plane model from {\sc Visilens}, and the red circle and point represent the critical and caustic lines; row 5 shows the source plane model with the black dashed ellipse showing the position of the source. All images are centered around the ALMA phase center (the (0.0, 0.0) coordinate in this case; see Table \ref{tab:lensing_params}).}
    \label{fig:img_model_lensing}
\end{figure*}

This appendix includes the figures detailing the {\sc Visilens} lensing model results, which are given in Fig.~\ref{fig:img_model_lensing}. 

\section{Lensing model for companions to BRI0952}

\begin{figure*}
    \centering
    \includegraphics[width = 0.8\linewidth]{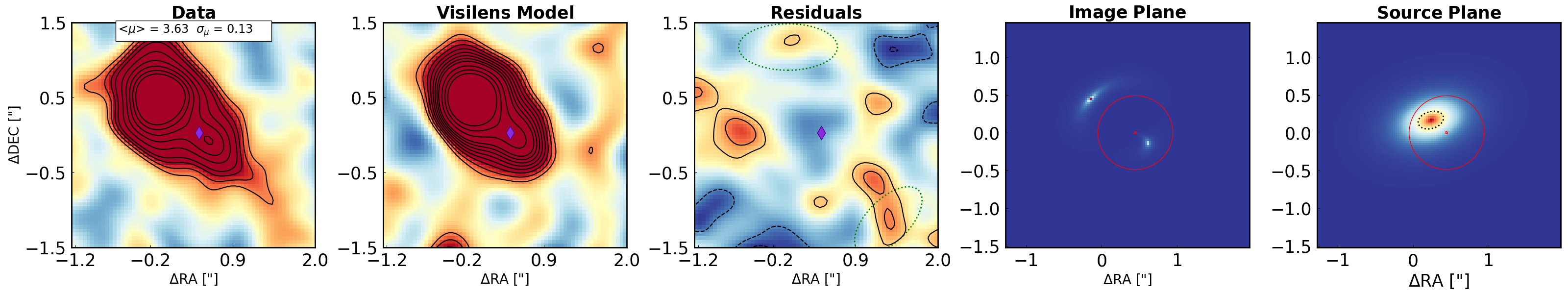}
    \caption{Image and lens models for BRI0952 isolated to the velocity range in which Comp-N and Comp-SW manifest. The diamond in the first three rows shows the best-fit position of the lens. Panels from top to bottom: Row 1 shows the ALMA band-7 dirty image with contours at 3, 4, 5, 6, 8, 10, 12, 14, and 16 $\sigma$ levels and the magnification value found for that bin; row 2 shows the image plane model from {\sc Visilens} with the same contour levels as row 1 and the residuals after the subtraction of the {\sc Visilens} model, contours are at -3, -2, 2, and 3 $\sigma$ levels; row 4 shows the image plane model from {\sc Visilens}, and the red circle and point represent the critical and caustic lines; row 5 shows the source plane model with the black dashed ellipse showing the position of the source. All images are centered around the ALMA phase center (the (0.0, 0.0) coordinate in this case; see Table \ref{tab:lensing_params}).}
    \label{fig:comp_lensing}
\end{figure*}

Appendix includes the figures detailing the {\sc Visilens} lensing model results for the companions of BRI0952, which are given in Fig.~\ref{fig:comp_lensing}. 

\section{BRI0952 Img-N \cii spectra}
Appendix includes the \cii spectra from Img-N of the quasar BRI0952 used in analysis of outflow properties as described in Section \ref{subsec:outflows}. 

\begin{figure}
    \centering
    \includegraphics[width = 0.99\linewidth]{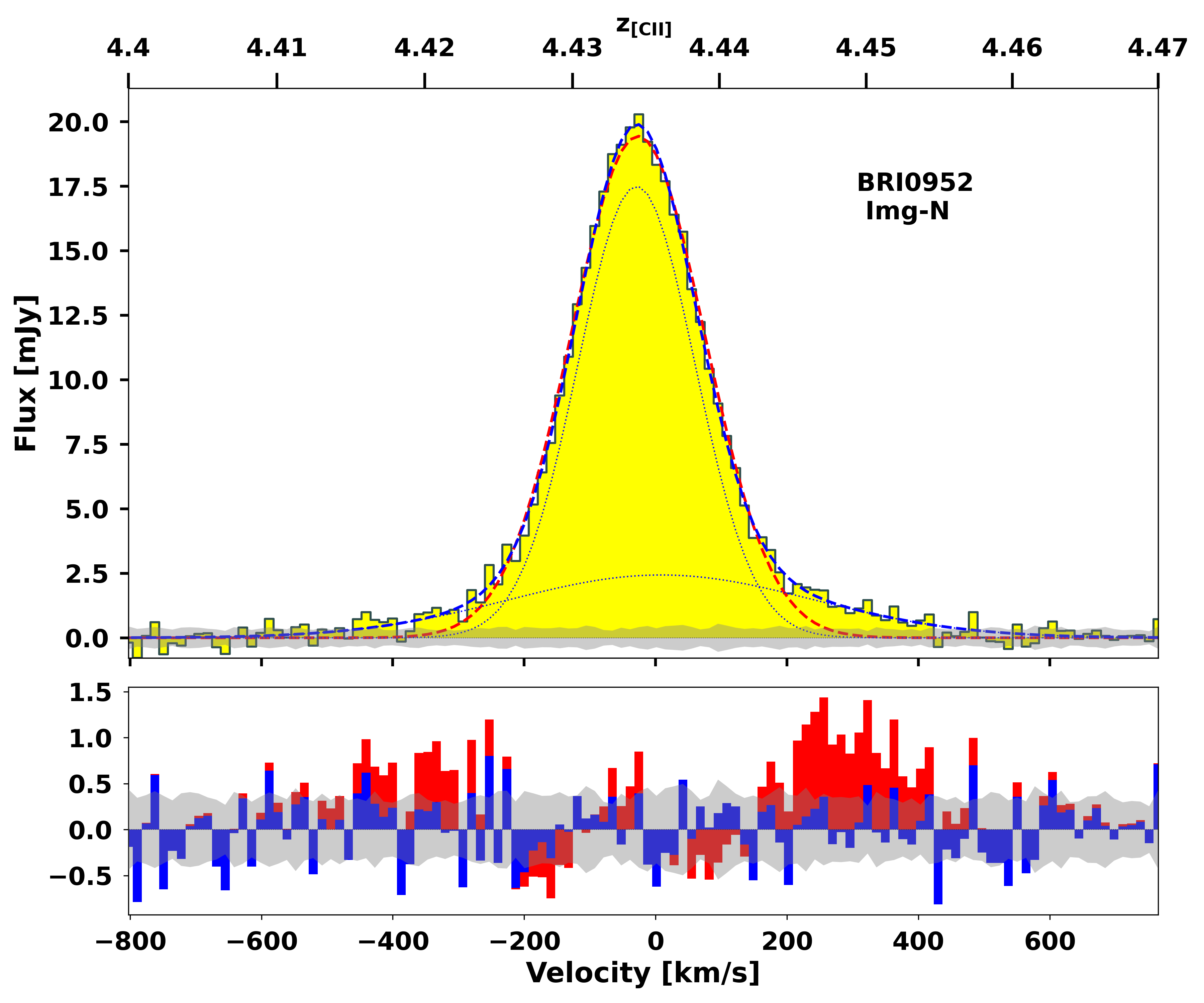}
    \caption{\cii spectra extracted from Img-N of the quasar BRI0952. The single Gaussian is plotted in red, whereas the double-Gaussian fit to the line is plotted in blue. The wings are clearly visible, from which the outflow rate was calculated when investigating the effect of differential magnification described in Section \ref{subsec:outflows}.}
    \label{fig:BRI_top_spec}
\end{figure}

\section{Velocity gradient of LBG-3}
This appendix includes a velocity map showing the velocity gradient detected in \cii between LBG-3 and AzTEC-3.
\begin{figure}
    \centering
    \includegraphics[width = 0.99\linewidth]{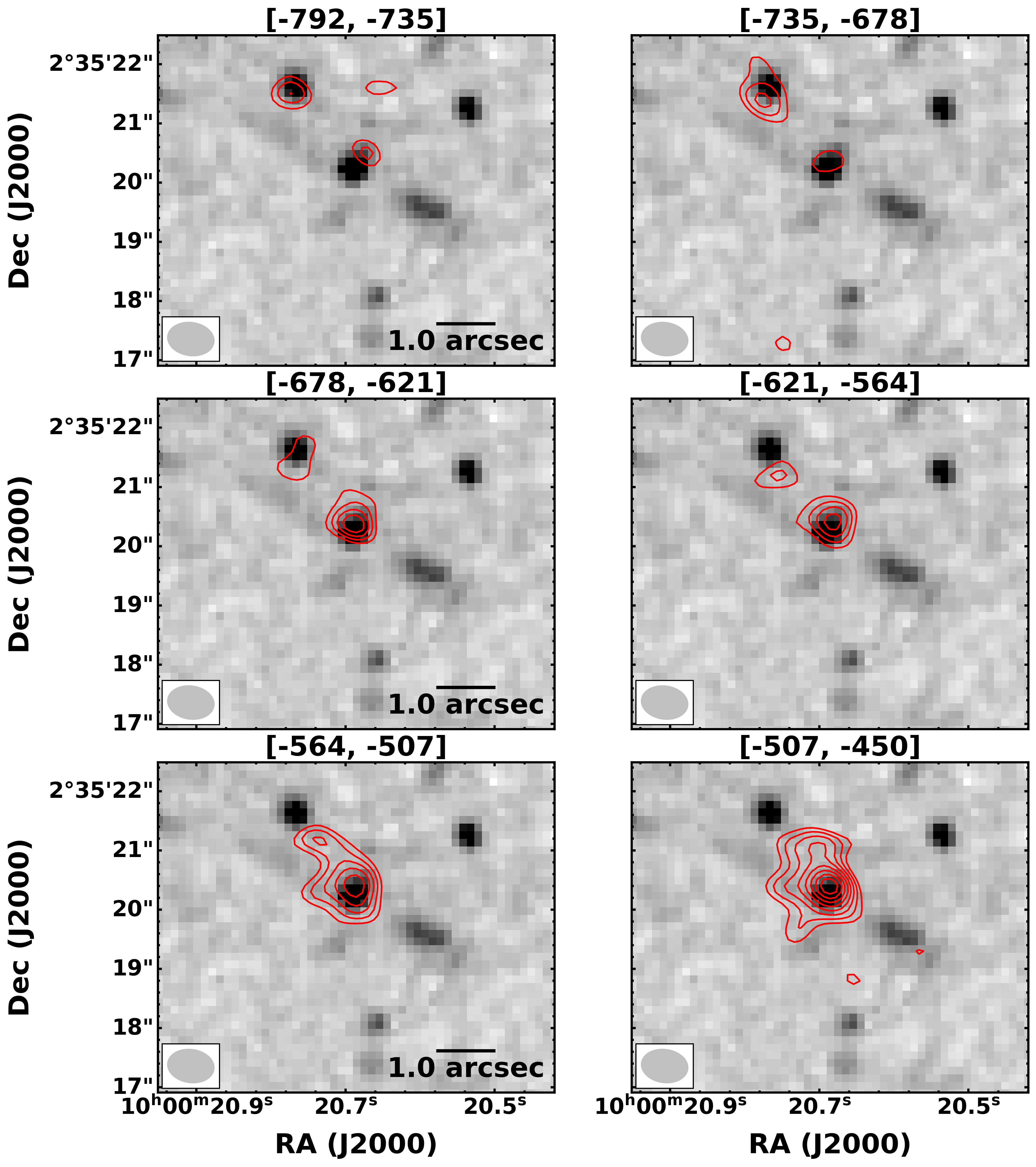}
    \caption{Velocity map of AzTEC-3 contours overlaid on \textit{HST} F105W centered around the velocities at which the slight velocity gradient between LBG-3 and AzTEC-3 is visible. The velocity range is added to the top of the images. The red contours are at 3, 4, 5, 6, 7, 8, 9, and 10 $\sigma$ levels.
    \label{fig:lbg3_velgrad}}
\end{figure}

\section{[CII] and continuum offset of LBG-3}
This appendix includes figures showing the \cii and continuum offset for LBG-3.

\begin{figure}
    \centering
    {\includegraphics[width = 0.99\linewidth]{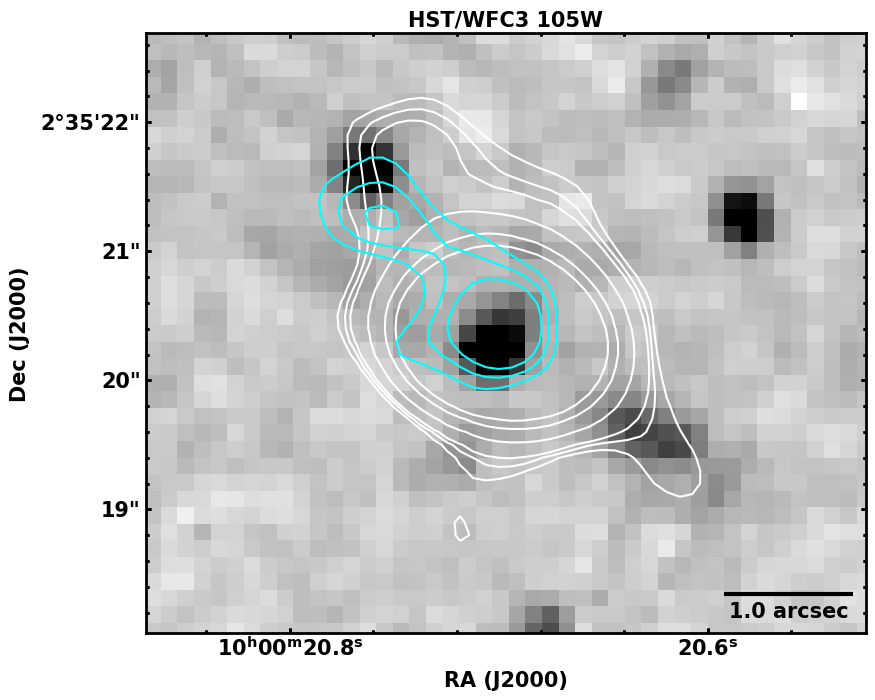}}
    \caption{\textit{HST} image of AzTEC-3 system with overlaid contours from the \cii emission for LBG-3 (cyan) and continuum contours (white). The \cii contours are shown at 3, 4, 5, and 6 $\sigma$ levels and 3, 4, 5, 6, 10, 20, and 30 $\sigma$ levels for the continuum. }
    \label{fig:cont_offset}
\end{figure}

\end{appendix}

\newpage

\end{document}